\RequirePackage[]{graphicx} 

\def\mode{1} 

\if 1\mode
	\documentclass[journal]{article}
\fi

\if 2\mode
	\documentclass[num-refs]{wiley-article_final}	
\fi

\usepackage{etoolbox}
\newtoggle{MRM}
\newtoggle{ARXIV}

\newtoggle{SUPMATERIAL}

\if 1\mode
	\toggletrue{ARXIV}
	\togglefalse{MRM}
\fi

\if 2\mode
	\toggletrue{MRM}
	\togglefalse{ARXIV}
\fi

\iftoggle{MRM}{
	\usepackage[nomarkers, nolists]{endfloat}
}

\newcommand{\Ncoil}{N_C}

\newcommand{\Nsampling}{N_S}

\newcommand{\NTrasp}{T_{\bm{R}}}
\newcommand{\NTcard}{T_{\bm{C}}}

\newcommand{\Ninv}{N_I}

\newcommand{\Nharm}{N_{H}}
\usepackage{amsmath}

\DeclareMathOperator*{\argmin}{arg\,min}

\usepackage[utf8]{inputenc}
\usepackage{amsmath}
\usepackage{bm}
\usepackage{bbold}
\usepackage{authblk}
\usepackage{amssymb}
\usepackage{graphicx}
\usepackage{ifdraft}
\usepackage{marginnote}
\usepackage{pdfpages}
\usepackage{tabularx,booktabs}
\usepackage{algorithm2e}
\usepackage{ifpdf}
\ifpdf
	\usepackage[]{hyperref}
\else
	\usepackage[dvipdfm]{hyperref}
\fi
\usepackage{tikz}
\usepackage[capitalise]{cleveref}

\usepackage{blindtext}
\usepackage{float}

\usepackage[
per-mode=fraction,
separate-uncertainty,
retain-unity-mantissa=false,
product-units=power,
table-alignment=center,
detect-all,
binary-units=true,
exponent-product=\cdot,
exponent-base=10
]{siunitx}
\usepackage{color}

\usepackage{changes}
\newcommand{\stkout}[1]{\ifmmode\text{\sout{\ensuremath{#1}}}\else\sout{#1}\fi}
\setdeletedmarkup{\stkout{#1}}

\title{\vspace{-3.5cm}Free-Breathing Myocardial $T_{1}$ Mapping using Inversion-Recovery Radial FLASH and Motion-Resolved Model-Based Reconstruction \footnote{Part of this work has been presented at the ISMRM Annual Conference 2021 (Virtual).}}

\title{{Free-Breathing Myocardial $T_{1}$ Mapping using Inversion-Recovery Radial FLASH and Motion-Resolved Model-Based Reconstruction}}
\newcommand{\authorA}{Xiaoqing Wang}
\newcommand{\authorB}{Sebastian Rosenzweig}
\newcommand{\authorC}{Volkert Roeloffs}
\newcommand{\authorD}{Moritz Blumenthal}
\newcommand{\authorE}{Nick Scholand}
\newcommand{\authorF}{Zhengguo Tan}
\newcommand{\authorG}{H. Christian M. Holme}
\newcommand{\authorH}{Christina Unterberg-Buchwald}
\newcommand{\authorI}{Rabea Hinkel}
\newcommand{\authorJ}{Martin Uecker}

\newcommand{\affilA}{Institute of Biomedical Imaging, Graz University of Technology, Graz, Austria}
\newcommand{\affilB}{Institute for Diagnostic and Interventional Radiology of the University Medical Center G\"ottingen, Germany}
\newcommand{\affilC}{German Centre for Cardiovascular Research (DZHK), Partner Site G\"ottingen, Germany}
\newcommand{\affilD}{Laboratory Animal Science Unit, Leibniz Institute for Primate Research, Deutsches Primatenzentrum GmbH, G\"ottingen, Germany}
\newcommand{\affilE}{Institute for Animal Hygiene, Animal Welfare and Farm Animal Behavior, University of Veterinary Medicine, Hannover, Germany}
\newcommand{\affilF}{Cluster of Excellence ``Multiscale Bioimaging: from Molecular Machines to Networks of Excitable Cells'' (MBExC), University of G\"ottingen, Germany}
\newcommand{\affilG}{BioTechMed-Graz, Graz, Austria}

\newcommand{\corAdress}{Xiaoqing Wang, Institute of Biomedical Imaging, Graz University of Technology, Stremayrgasse 16/III, 8010 Graz, Austria.}

\newcommand{\corMail}{ xiaoqingwang2010@gmail.com}
\newcommand{\runHead}{Motion-resolved model-based reconstruction for free-breathing myocardial $T_{1}$ mapping}

\iftoggle{ARXIV}{
		\author[1,2,3]{\authorA \thanks{\corAdress \corMail}}
	}

\iftoggle{MRM}{
		\author[1,2]{\authorA}
	}
\author[2,3]{\authorB}
\author[2]{\authorC}
\author[1,2]{\authorD}
\author[1,3]{\authorE}
\author[2,3]{\authorF}
\author[1]{\authorG}
\author[2,3]{\authorH}
\author[3,4,5]{\authorI}
\author[1,2,3,6,7]{\authorJ}

\affil[1]{\affilA}
\affil[2]{\affilB}
\affil[3]{\affilC}
\affil[4]{\affilD}
\affil[5]{\affilE}
\affil[6]{\affilF}
\affil[7]{\affilG}
	
\iftoggle{MRM}{
	\papertype{Note}
	\corraddress{\corAdress}
	\corremail{\corMail\\
			
			Submitted to XXX
			\bf{{Word count}}: Abstract , Body .}
	
	\runningauthor{\runHead}
}

\begin{document}
	
\maketitle

\begin{abstract}
\textbf{Purpose}: To develop a free-breathing myocardial $T_{1}$ mapping technique 
using inversion-recovery (IR) radial fast low-angle shot (FLASH) 
and calibrationless motion-resolved model-based reconstruction.

\noindent \textbf{Methods}: Free-running (free-breathing, retrospective cardiac gating) IR radial FLASH is used for
data acquisition at 3T. First, to reduce the waiting time between 
inversions, an analytical formula is derived that takes the
incomplete $T_{1}$ recovery into account for an
accurate $T_{1}$ calculation. Second, the respiratory
motion signal is estimated from the k-space center of 
the contrast varying acquisition
using an adapted singular spectrum analysis (SSA-FARY) technique. 
Third, a motion-resolved model-based reconstruction is
used to estimate both parameter and coil sensitivity maps
directly from the sorted k-space data.  Thus,
spatio-temporal total variation, in addition to the spatial 
sparsity constraints, can be directly applied to the parameter 
maps.  Validations are 
performed on an experimental phantom, eleven human subjects,
and a young landrace pig with myocardial infarction.

\noindent
\textbf{Results}: In comparison to an IR spin-echo reference, phantom results confirm good $T_{1}$ accuracy, when reducing the waiting time from five seconds to one second using the new correction.
The motion-resolved model-based reconstruction further improves $T_{1}$ precision compared to the spatial regularization-only reconstruction. 
Aside from showing that a reliable respiratory motion signal can be estimated using
modified SSA-FARY, in vivo studies demonstrate that dynamic myocardial $T_{1}$ maps 
can be obtained within two minutes with good precision and repeatability.

\noindent \textbf{Conclusion}: 
Motion-resolved myocardial $T_{1}$ mapping during free-breathing
with good accuracy, precision and repeatability can be
achieved by combining inversion-recovery
radial FLASH, self-gating and a calibrationless motion-resolved model-based reconstruction.

\noindent
\textbf{Keywords}: free-breathing myocardial $T_{1}$ mapping, self-gating, motion-resolved model-based 
reconstruction, radial FLASH, spatiotemporal total variation
\end{abstract}

\section{Introduction}

Quantitative myocardial $T_{1}$ mapping is becoming ever more important in clinical 
cardiovascular magnetic resonance (CMR) imaging \cite{Moon_J.Cardiovasc.Magn.Reson._2013,
	Kellman_J.Cardiovasc.Magn.Reson._2014}. For example, both native and post-contrast 
$T_{1}$ mapping can be used to assess
diffuse myocardial fibrosis \cite{Puntmann_JACC:Cardiovascular.Imaging_2013}. Commonly used $T_{1}$ mapping 
techniques are modified Look-Locker inversion recovery (MOLLI) \cite{Messroghli_Magn.Reson.Med._2004}, 
saturation recovery single-shot acquisition (SASHA) \cite{Chow_Magn.Reson.Med._2014},
and saturation pulse prepared heart rate independent inversion recovery (SAPPHIRE) \cite{Weingartner_Magn.Reson.Med._2014}.
These techniques normally utilize a breathhold to mitigate respiratory motion and 
use an external electrocardiogram (ECG) device to synchronize data acquisition 
to a certain cardiac phase (e.g., end-diastolic), reducing the influence of cardiac motion.
Although widely used, the need of a breathhold time of around 11 to 17 heartbeats may cause discomfort
for patients (such as heart failure patients) and limits the achievable spatial resolution.
Substantial efforts were made to shorten the 
breathhold period by optimized sampling \cite{Piechnik_J.Cardiov.Magn.Reson._2010}, 
or by using non-Cartesian acquisition for single-shot myocardial $T_{1}$ mapping \cite{Gensler_Radiology_2014, 
	Wang_Brit.J.Radiol._2016,  Marty_Magn.Reson.Med._2018} or by cardiac magnetic resonance fingerprinting (MRF) techniques for efficient multi-parameter mapping \cite{Hamilton_Magn.Reson.Med._2017,Jaubert_J.Magn.Reson.Imaging_2021, liu_Magn.Reson.Med._2021, Lima_Magn.Reson.Med.2022}. 
More recently, free-breathing strategies 
\cite{Christodoulou_NatureBiomedicalEngineering_2018, Shaw_Magn.Reson.Med._2019,
	Qi_Magn.Reson.Med._2019, Guo_Magn.Reson.Med._2021, Zhou_Magn.Reson.Med._2021} were investigated.
These approaches acquire data continuously without the need for breath-holding and extract 
motion (respiration and/or cardiac) signals from the measured data itself using self-gating techniques. 
Following motion-resolved image 
reconstruction and pixel-wise fitting/matching, cardiac $T_{1}$ maps can then be obtained for certain motion states. 

Model-based reconstruction \cite{Block_IEEETrans.Med.Imaging_2009,
	Fessler_IEEESignalProcess.Mag._2010, Wang_Philos.Trans.R.Soc.A._2021} 
is an alternative approach to quantitative MRI. These methods
directly reconstruct parameter maps from k-space, substantially reducing the 
number of unknowns to the number of actual physical parameters
by not first reconstructing contrast-weighted images.
They also offer a flexible choice of temporal footprint for parameter quantification 
as no intermediate image reconstruction is needed. Furthermore, sparsity constraints 
can be applied directly to the parameter maps to improve
precision \cite{Zhao_IEEETrans.Med.Imag._2014, Wang_Magn.Reson.Med._2018, 
	Maier_Magn.Reson.Med._2018}. Model-based approaches have been used to accelerate 
myocardial $T_{1}$ mapping at high spatial resolution \cite{Becker_Magn.Reson.Med._2019, 
	Wang_J.Cardiovasc.Magn.Reson._2019}, but still require breath-holding.

Combining idea from all these strategies, we aim to develop a free-breathing 
myocardial $T_{1}$ mapping technique by combining a free-running inversion-prepared radial FLASH 
sequence, an adapted self-gating technique and a calibrationless motion-resolved model-based 
reconstruction. In particular, the techniques integrate three novel developments:
First, instead of setting the delay long enough to 
allow for a full $T_{1}$ recovery ($>$ 5 s), we have derived an analytical formula for
accurate $T_{1}$ calculation even when $T_{1}$ recovery is incomplete, i.e., $\leq
$ 3 s. Second, to allow for robust respiratory
motion estimation, we propose to use an extended technique based on
SSA-FARY \cite{Rosenzweig_IEEETrans.Med.Imag._2020} to extract the respiratory motion signal from the k-space center by eliminating the trajectory-dependent oscillations
and inversion contrast in a preprocessing step. Third, after sorting
raw data into a number of respiration and cardiac bins based on the 
estimated respiration signal and the recorded ECG signal, we estimate both parameter
maps and  coil sensitivity maps of the desired motion bins directly from k-space using a calibrationless 
motion-resolved model-based reconstruction. The latter is an extension of a previously developed
model-based reconstruction \cite{Wang_Magn.Reson.Med._2018, 
	Wang_J.Cardiovasc.Magn.Reson._2019} to the motion-resolved case which enables the application of sparsity 
constraints along all motion dimensions, in addition to the spatial regularization, 
to further improve $T_{1}$ precision. 
Validation of the proposed method was performed on an experimental phantom, eleven healthy subjects and one landrace pig with infarcted myocardium. 

\section{Theory}

\label{sec:theory}
\subsection*{Sequence Design and $T_{1}$ Estimation from Incomplete Recovery}
\label{subsec:sequence}
The free-running $T_{1}$ mapping sequence is shown in Figure~1 (A). 
It consists of three repeated blocks: (1) non-selective inversion 
(2) a continuous radial FLASH readout using a tiny golden-angle \mbox{($\approx23.63^\circ$)} \cite{Wundrak_Magn.Reson.Med._2016_2} with a 3-s
duration (3) and a time delay ($T_{1}$ recovery) before the inversion in the next repetition. In previous studies 
using multiple inversions \cite{Wang_Magn.Reson.Med._2020, Feng_Magn.Reson.Med._2021}, the delay time was set long
enough to ensure a full recovery of longitudinal magnetization so that $T_{1}$ can still be calculated using the 
conventional Look-Locker formula \cite{Look_Rev.Sci.Instrum._1970, Deichmann_J.Magn.Reson._1992}. 
However, a full recovery may need as long as 5 seconds for cardiac 
$T_{1}$ mapping, which prolongs the total acquisition time. In this work, we treat this delay as a period 
that encodes $T_{1}$ information in the data. We use an analytical formula based on $T_{1}$, $T_{1}$*, 
the steady-state signal $M_{ss}$, and the new start magnetization signal $M_{0}^{'}$ (i.e., in the 
case that the $T_{1}$ recovery is
not complete: $M_{0}^{'} < M_{0}$, with $M_{0}$ the equilibrium magnetization):
\begin{equation}
\label{T1_est_parRev}
M_{0}' = \frac{R_{1}M_{ss}R_{1}^{*}(1-E_{1})+ E_{1} M_{ss} (1-E_{1}^{*})}{1 + E_{1}\cdot E_{1}^{*}}
\end{equation}
where $E_{1} = e^{-R_{1}\cdot t_{1}}$, $E_{1}^{*} = e^{-R_{1}^{*}\cdot t_{1s}}$, $R_{1} = 1/T_{1}$, 
$R_{1}^{*} = 1/T_{1}^{*}$ and $t_{1}$, $t_{1s}$ are the time periods 
for $T_{1}$ and $T_{1}$* relaxation, respectively. Thus, $T_{1}$ can be estimated even from partial $T_{1}$ 
recoveries after reconstruction of the parameter maps $(M_{ss}, M_{0}^{'}, R_{1}^{*})^{T}$ according to Equation~(\ref{T1_est_parRev}).
Here, we adopt a bisection root-finding algorithm to solve Equation~ (\ref{T1_est_parRev}).
A full derivation of the above equation can be found in the Supporting Information File.

\subsection*{Respiratory Motion Estimation}
\label{subsec:self-gating}
	The main steps of the respiratory motion estimation process are demonstrated in the flowchart in Figure~1 (B). In the following, we explain all these steps in detail.
Similar to \cite{Larson_Magn.Reson.Med._2004, Rosenzweig_IEEETrans.Med.Imag._2020}, we 
construct an auto-calibration (AC) region for self-gating using the central k-space samples 
of a radial acquisition, resulting in a time-series $X(t)$ of size 	
$[\Ncoil \times N_{t}]$,
with $\Ncoil$ and $N_{t}$ the total number of channels (phased array coils) and central k-space points, respectively. 
$N_{t}=\Nsampling \cdot \Ninv$, with $\Nsampling$ 
the number of sampling points per IR and
$\Ninv$ the total number of inversions. 
The AC data is usually corrupted by a trajectory-dependent signal 
due to eddy currents. Therefore, we first remove such oscillations 
by extending the method of
orthogonal projections (with higher-order harmonics) from the steady-state case 
\cite{Rosenzweig_IEEETrans.Med.Imag._2020} 
to the contrast-change case (inversion recovery). 
Details of this procedure can be found
in the Section II of the Supporting Information File.

Following removal of the oscillations, the new k-space center 
signal $\tilde X(t)$ 
can be modeled as 
\begin{equation}
\label{self-gating_signal}
\tilde{X}(t) = s(t)\cdot m_{1}(t) + m_{2}(t)
\end{equation}
with $s(t)$ the steady-state signal which contains motion information (ideally without contrast change),
and $m_{1}(t)$ and $m_{2}(t)$ the multiplicative and additive 
signals which model the contrast change due to inversion \cite{Winter_Magn.Reson.Med._2016}. 
Here we propose the following procedure to 
remove the main effects from the changing contrast 
and to extract the signal component that is most 
relevant for respiratory motion:
\begin{itemize}
	\item Step 1. Estimating $m_{2}(t)$: Perform the singular spectrum analysis (SSA) 
	on $\tilde{X}$(t), remove the components that are mostly related to the inversion-recovery contrast in the spectrum domain 
	and transfer the processed signal back to the time domain. In SSA \cite{Rosenzweig_IEEETrans.Med.Imag._2020}, this step 
	largely removes the additive contrast-changing component of Equation~(\ref{self-gating_signal}), resulting in
	a new signal $\tilde{X_{1}}(t) = \tilde{X}(t) - \tilde{m}_{2}(t)$, with $\tilde{m}_{2}(t)$
	the estimated additive contrast-changing signal.
	
	\item Step 2. Estimating $m_{1}(t)$: Since the multiplicative component is mainly left in 
	Equation~(\ref{self-gating_signal}), the singular value decomposition (SVD) is then performed on $\tilde{X_{1}}(t)$ and the corresponding 
	rank-one approximation is taken, generating an estimate of the multiplicative component $\tilde{m}_{1}(t)$.  
	Next, the magnitude of $\tilde{m}_{1}(t)$ was utilized for the calculation, leading to a new signal $\tilde{s}(t) = \frac{\tilde{X_{1}}(t)}{|\tilde{m}_{1}(t)|}$. Due to the $180^{\circ}$ phase difference before and after zero-crossing caused by inversion, the phase
	of $\tilde{s}(t)$ before zero-crossing needs to be inverted. This was done by first detecting the 
	minimum points of the absolute value of signal $\tilde{X}$(t) (zero-crossing) for each coil 
	and inversion and then correcting using:
	\begin{equation}
	\label{eq_zc}
	\tilde{s}_{1}(t) =
	\begin{cases}
	-\tilde{s}(t),& \text{if } t \leq \text{zero-crossing}\\
	\tilde{s}(t),              & \text{otherwise}.
	\end{cases}
	\end{equation} 	 
	\item Step 3. Zero-padding: 
	To account for the missing temporal information due to the delay time between inversions, $\tilde{s}_{1}(t)$
	was zero-padded, 
	generating a new signal $\tilde{s}_{2}(t) \in 	
	C^{[\Ncoil \times (\Ninv \cdot (\Nsampling + N_{\text{Z}}))]}$ with $N_{\text{Z}}$
	calculated by $N_{\text{Z}}$ = $\frac{\text{Delay Time}}{\text{Repetition Time (TR)}}$.
	
	\item Step 4. SSA-FARY: Perform SSA-FARY on $\tilde{s}_{2}(t)$ with the window size tuned to estimate
	the signal component that is most relevant for the respiratory motion.

\end{itemize}

\subsection*{Motion-Resolved Model-Based Reconstruction}
The acquired k-space data is then sorted into 6 respiration and 20 cardiac bins using amplitude 
binning \cite{Rosenzweig__2021a} based on the estimated respiratory signal and the recorded 
ECG signal. The MR physical parameter maps in Equation~(\ref{T1_est_parRev}) for the selected motion 
states are estimated directly from k-space using a calibrationless model-based 
reconstruction \cite{Wang_Magn.Reson.Med._2018, Wang_J.Cardiovasc.Magn.Reson._2019}. Here, to further 
exploit sparsity along the motion dimensions, the previous model-based reconstruction is extended to 
the motion-resolved case by formulating the estimation of unknowns from the selected motion states as a single regularized nonlinear inverse problem:
\begin{equation}
\label{eq_moba}
\hat{x} = \argmin_{x\in S} \sum_{r=1}^{\NTrasp}\sum_{c=1}^{\NTcard}\Big\|F_{r,c}(x) -{Y}_{r,c} \Big\|_{2}^{2} 
+ \alpha R(x_{\bm{m}}) + \beta U(x_{\bm{c}})
\end{equation}
where $F$ is a nonlinear operator \cite{Wang_J.Cardiovasc.Magn.Reson._2019} mapping all 
unknowns $x$ to	the sorted k-space data ${Y}$. $\NTrasp$, $\NTcard$ are 
the numbers of respiration and cardiac bins, respectively. 
$x = (x_{\bm{m}}, x_{\bm{c}})^{T}$ where $x_{\bm{m}}$ contains MR physical parameter maps in \mbox{Equation~(\ref{T1_est_parRev})}, i.e., 
$(M_{ss}, M_{0}^{'}, R_{1}^{*})^{T}$ of all the selected motion states and $x_{\bm{c}}$ 
represents coil sensitivity maps $(c_{1}, \cdots, c_{N_{\text{C}}})^{T}$ for the corresponding motion states. 
$S$ is a convex set ensuring non-negativity of the relaxation rate ${R^{*}_{1}}$.
For the regularization $R(\cdot)$, we first adopt the joint $\ell_{1}$-Wavelet spatial constraints \cite{Wang_Magn.Reson.Med._2018}. Second, 
we add the temporal total variation (TV) regularization to explore sparsity along the motion dimensions \cite{Feng_Magn.Reson.Med._2016a}. Furthermore, as a pure TV
may favor straight lines if applied along the motion dimension only, we utilize 
a joint TV regularization along spatial and temporal dimensions to better preserve the spatio-temporal 
information. 
Thus, $R(\cdot)$ reads: 
\begin{equation}
\label{eq_2}
R(x_{\bm{m}}) = \lambda_{1}\|Wx_{\bm{m}}\|_1 + \sqrt{\lambda_{2} \|D_{x} x_{\bm{m}}\|^{2} +
	\lambda_{3} \|D_{y} x_{\bm{m}}\|^{2} +
	\lambda_{4} \|D_{c} x_{\bm{m}}\|^{2} +
	\lambda_{5} \|D_{r} x_{\bm{m}}\|^{2}}
\end{equation}
with $\|Wx_{\bm{m}}\|_1$ the joint $\ell_{1}$-Wavelet spatial regularization and $D_{x}$, $D_{y}$, $D_{c}$ and $D_{r}$ the 
gradient operators along the $x$, $y$, cardiac and respiratory dimensions, respectively. 
$\lambda_{1}$, $\lambda_{2}$, $\lambda_{3}$, $\lambda_{4}$ and $\lambda_{5}$ are the corresponding 
weighting parameters, balancing the effects of different regularization terms.
$\alpha$ is a global regularization parameter on $R(\cdot)$. 
$U(\cdot)$ represents the Sobolev regularization term \cite{Uecker_Magn.Reson.Med._2008} on the coil 
sensitivity maps with $\beta$ the regularization parameter. Similar to 
\cite{Wang_Magn.Reson.Med._2018, Wang_J.Cardiovasc.Magn.Reson._2019}, the above nonlinear 
inverse problem is solved by the iteratively regularized Gauss‐Newton method (IRGNM) 
algorithm \cite{Bakushinsky__2005} where the nonlinear
problem in Equation~(\ref{eq_moba}) is linearizedly solved in each Gauss-Newton step. To enable the use of multiple regularizations, the ADMM algorithm 
\cite{Boyd_Found.TrendsMach.Learn._2011} was employed to solve the linearized subproblem. More details of the proposed IRGNM-ADMM algorithm can be found in the Appendix.

\section{Methods}

\subsection*{Data Acquisition}
All MRI experiments were performed on a Magnetom Skyra 3T (Siemens Healthineers, 
Erlangen, Germany) with approval of the local ethics committee. Animal care and all experimental procedures were performed in strict accordance with the German and National Institutes of Health animal legislation guidelines and were approved by the local animal care and use committees. Validations were
first performed on a commercial reference
phantom (Diagnostic Sonar LTD, Scotland, UK) consisting of 6 compartments with defined
$T_{1}$ values surrounded by water. Phantom scans employed a 20-channel head/neck coil, 
while in vivo measurements used combined thorax and spine coils with 26 channels.  
In the phantom study, the delay time TD in the sequence was varied from 5 s to 1 s (with a step size of 1 s) 
to study $T_{1}$ accuracy and precision when using the proposed $T_{1}$ estimation procedure. 
An optimal value of TD was then chosen for subsequent in vivo studies.
Informed written consent was obtained from all subjects prior to MRI. In vivo scans were performed
during free-breathing using the free-running sequence. The ECG signals were recorded for later use but not for triggering. To assess 
repeatability of the proposed method, the sequence was repeated twice in the middle short-axis slice for 
each subject. Data from basal and apical slices were additionally acquired for a subset of subjects.  
After excluding measurements with non-reliable ECG signal, data sets from eleven subjects 
(seven female, four male, age 25 $\pm$ 4, range 21 - 37 years; 
heart rates 63 $\pm$ 9 bpm, range 50 - 77 bpm) were used in this work (including six subjects
with basal and apical slices).
For FLASH readout, spoiling of the residual transverse magnetization was achieved using
random radiofrequency (RF) phases \cite{Roeloffs_Magn.Reson.Med._2016}. The other acquisition parameters
were: field of view (FOV) = $256 \times 256$ mm$^{2}$, slice thickness = 6 mm, 
matrix size = $256 \times 256$, TR/echo time (TE) = 3.27-3.30/1.98 ms, RF-pulse time-bandwidth product = 4.5, 
nominal flip angle = $6^{\circ}$, receiver bandwidth = 810 Hz/pixel and
total acquisition time of 21 inversions, i.e., around 2 min with 915 radial spokes per inversion. 
A window size of 21 frames 
was chosen for the adapted SSA-FARY technique \cite{Rosenzweig_IEEETrans.Med.Imag._2020}. 
Further, one data set from a young landrace
pig with infarcted myocardium (regions in the septum and anterior wall due to
intermittent left anterior descending artery occlusion) was acquired on the mid-ventricular
short-axis slice using the same free-running acquisition parameters described above.

For reference, gold standard $T_{1}$ mapping was performed on the phantom using
an IR spin-echo method \cite{Barral_Magn.Reson.Med._2010} with 9 IR scans
(TI = 30, 530, 1030, 1530, 2030, 2530, 3030, 3530, 4030 ms), 
TR/TE = 4050/12 ms, FOV = 192 $\times$ 192 $\text{mm}^{2}$, slice thickness = 5 mm,
matrix size = 192 $\times$ 192, and a total acquisition time of 2.4 h.
For in vivo studies, a 5(3)3 MOLLI reference sequence with a hyperbolic tangent inversion pulse \cite{Kellman_Magn.Reson.Med._2014} provided by the vendor was applied 
for end-diastolic $T_{1}$ mapping using
a FOV of $360 \times 306.6$ $\text{mm}^{2}$, 
in-plane resolution = $1.41\times1.41\times8$ $\text{mm}^{3}$, TR/TE = 2.7/1.12 ms,
nominal flip angle = $35^\circ$, receiver bandwidth= 1085 Hz/pixel and an 
acquisition period of 11 heart beats during a single breathhold. A correction factor was further applied to the final MOLLI $T_{1}$ map to accommodate for the imperfect inversion \cite{Kellman_Magn.Reson.Med._2014}.


\subsection*{Iterative Reconstruction}

The motion-resolved model-based reconstruction algorithm was implemented using the nonlinear operator and 
optimization framework in C/CUDA in the Berkeley Advanced Reconstruction Toolbox (BART) \cite{Uecker__2015}. To reduce 
computational demand, we selected three respiratory motion bins (out of 6) close to the end-respiratory state and 
all cardiac bins for the motion-resolved quantitative reconstruction.
Similar to \cite{Wang_Magn.Reson.Med._2018}, we initialized the parameter maps $(M_{ss}, M^{'}_{0},
{R^{*}_{1}})^{T}$ with $(1.0, 1.0, 1.5)^{T}$ and all coil
sensitivities with zeros in the IRGNM-ADMM algorithm. 
Moreover, as a high accuracy is usually not necessary during 
the first Gauss-Newton steps, 
we set the number of ADMM iteration steps to be $N_{n} = \min(100, 2N_{n-1})$ 
at the $n$-th Gauss-Newton step with $N_{0} = 10$. This setting resulted in stable reconstructions 
for all cases tested.

Regularization parameters were tuned to balance the preservation of  
image details versus reduction of noise. 	
The regularization parameters $\alpha$ and $\beta$ were initialized with $1.0$ and
subsequently reduced by a
factor of three in each Gauss-Newton step. A minimum value of $\alpha$ was used to
control the noise of the estimated parameter maps even with a large number
of Gauss-Newton steps, i.e., $\alpha_{n+1} = \text{max}\big(\alpha_{\text{min}}, (1/3)^{n}\cdot \alpha_{0}\big)$. 
The optimal value $\alpha_{\rm min}$ as well as the weighting parameters $\lambda$s were chosen 
manually to optimize the signal-to-noise ratio (SNR) without compromising the quantitative accuracy or delineation of
structural details. Particularly, $\alpha_{\rm min}$ was tuned from 0.004 to 0.007 with the optimal value chosen by visual inspection.  $\lambda_{1}$ was 
set to be 0.2 and parameters $(\lambda_{2}, \lambda_{3}, \lambda_{4}, \lambda_{5})^{T}$ in the weighted
spatio-temporal TV regularization term in Equation~(\ref{eq_2}) were set to be $(0.4, 0.4, 1.0, 0.2)^{T}$. 
For comparison, the other types of regularization, such as the spatial-only
($\ell_{1}$-Wavelet) regularization ($R(x_{\bm{m}}) = \lambda_{1}\|Wx_{\bm{m}}\|_{1}$), temporal TV regularization 
($R(x_{\bm{m}}) = \lambda_{4}\|D_{c}x_{\bm{m}}\|_{1} + \lambda_{5}\|D_{r}x_{\bm{m}}\|_{1}$) and the combination of 
the above two ($R(x_{\bm{m}}) = \lambda_{1}\|Wx_{\bm{m}}\|_{1} + \lambda_{4}\|D_{c}x_{\bm{m}}\|_{1} + \lambda_{5}\|D_{r}x_{\bm{m}}\|_{1}$) 
were also implemented and first evaluated on a simulated dynamic phantom 
using the parallel imaging and compressed sensing (PICS) tool in BART, followed by the evaluation on the data
of one subject with the proposed motion-resolved model-based reconstruction. More details regarding the simulated dynamic 
phantom can be found in Section III of the Supporting Information File.

With the above parameter settings, all image reconstruction was done offline.
After gradient-delay
correction \cite{Rosenzweig_Magn.Reson.Med._2018a} and
channel compression to six principal components, the multi-coil radial data
were gridded onto a Cartesian grid, where all successive iterations were
then performed using FFT-based convolutions with the point-spread function
\cite{Uecker_Magn.Reson.Med._2010}. To reduce memory demand during iterations, 15 spokes were 
binned into one k-space frame 
prior to model-based reconstruction, resulting in a nominal temporal resolution of around 49 ms. 
To allow for efficient reconstructions, implementations were optimized in BART (see Section IV of the Supporting Information File) so that all 
computations could run on a GPU (A100, NVIDIA, Santa Clara, CA) with a memory of 80 GB.
It then took around
20 - 30 minutes to reconstruct one in vivo data set using the above reconstruction parameters.

\subsection*{$T_{1}$ Analysis}

All quantitative $T_{1}$ results are reported as mean $\pm$ standard deviation (SD). For the in vivo studies, 
$T_{1}$ maps from the end-diastolic and end-respiratory phase were selected for quantitative assessment 
of the proposed method. 
Regions-of-interest (ROIs) were carefully drawn into the myocardial segments model defined by 
the American Heart Association (AHA) \cite{Pearson_Circulation_2002} with 6 segments in the basal 
and middle slices and 4 segments in the apical slice using the arrayShow \cite{Sumpf__2013} tool in
MATLAB (MathWorks, Natick, MA). The mean $T_{1}$ values were calculated for each segment across all 
subjects and scans, and were visualized with bullseye plots. The repeatability error was calculated using 
$\sqrt{\big(\sum_{i=1}^{n_{s}}\text{$T_{1}$}^{2}_{\text{diff}}(i)\big)/n_{s}}$, 
with $\text{$T_{1}$}_{\text{diff}}(i)$ 
the $T_{1}$ difference between two repeated measurements and $n_{s}$ the number of subjects. 
The precision of $T_{1}$ estimation was computed using the coefficient of variation 
(CoV = $\text{SD}_\text{ROI}$ / $\text{Mean}_\text{ROI}$ $\times$ 100\%). Further, Bland-Altman analyses 
were performed to compare ROI-based mean $T_{1}$ values between different $T_{1}$ mapping techniques. The two-tailed 
Student's t-tests were utilized for comparison, and a $p$ value $<$ 0.05 was considered significant. In addition, to quantify
the in-plane cardiac motion between end-diastolic and end-systolic phases, the relative difference of the left-ventricular 
area (($\text{Area}_{\text{End-Diastolic}} - \text{Area}_\text{End-Systolic}$) / ${\text{Area}_\text{End-Diastolic}}$  $\times$ 100\%), 
similar to the left-ventricular ejection fraction (LVEF) index for the volume case, was calculated based on the mid-ventricular 
myocardial $T_{1}$ maps. The blood–myocardium boundary was manually segmented using the aforementioned arrayShow tool.
\section{Results}
	\subsection*{Phantom Validation}
We first validated the proposed $T_{1}$ correction procedure for phantom $T_{1}$ mapping when
using different delay times in the multi-shot acquisition in comparison to an IR spin-echo reference.
Figure~2 presents the estimated $T_{1}$ maps for acquisitions with 
delay times ranging from five seconds to one second (step size one second)
when using the conventional Look-Locker formula and the proposed procedure. 
Prior to $T_{1}$ correction, all the physical parameters $(M_{ss}, M_{0}^{'}, R_{1}^{*})^{T}$ were 
estimated by the single-slice model-based reconstruction \cite{Wang_Magn.Reson.Med._2018} using the
data from the second inversion, where the initial magnetization is affected by incomplete recovery. 
Both quantitative $T_{1}$ maps and $T_{1}$ values of a ROI in Figure~2 as well as the Bland-Altman plots 
in the Supporting Information Figure~S1 (A) reveal that the conventional Look-Locker correction 
underestimates $T_{1}$: the smaller the delay time, the higher the 
bias. On the other hand, the proposed procedure could achieve good $T_{1}$ accuracy 
regardless of the delay time, but at the expense of increased noise (i.e., lower precision) for short delays
and large $T_1$ times. 
This is mainly due to the fact that there is less $T_{1}$ information encoded in the data for shorter delays. 
In the extreme case where there is no delay, it is impossible to recover $T_{1}$ (i.e., decouple 
$T_{1}$ and $B_{1}^{+}$ from $T_{1}^{*}$) as no explicit $T_{1}$ is 
encoded in the data. According to these results, a delay time of two or three seconds is a good compromise 
between short acquisition time and good $T_{1}$ precision. We choose three seconds for the other acquisitions in this study. 

Subsequently, we evaluated the proposed motion-resolved model-based reconstruction on the same phantom using the 
multi-shot data with the delay time of three seconds. The data has been sorted into 6 respiratory and 20 cardiac motion states 
based on the motion signals estimated from one human subject (subject $\#$3, scan $\#$1). Figure~3 (top) shows
the estimated phantom $T_{1}$ maps (selected at the end-expiration and end-diastolic phase) with the spatial-only regularization 
using two different regularization parameters $\alpha_{\min} = 0.005$ and $\alpha_{\min} = 0.02$, 
and its combination with the spatio-temporal TV regularization with $\alpha_{\min} = 0.005$. Figure~3 (bottom) plots 
the corresponding quantitative $T_{1}$ values of the ROI against the IR spin-echo reference. 
The Supporting Information Figure~S1 (B) further presents the corresponding Bland-Altman plots. The above
quantitative results show that all reconstructions could achieve good $T_{1}$ accuracy. 
The increase of the regularization strength in the spatial-only regularization or the use of an additional 
spatio-temporal TV with the same regularization parameter is helpful for reducing noise (improving $T_{1}$ precision) 
in the quantitative phantom $T_{1}$ maps.

\subsection*{In Vivo Studies}
\subsubsection*{Respiratory Motion Estimation}
Figure~4 shows the DC component for one inversion recovery before and after data correction using 
the extended orthogonal projection with the order of harmonics $\Nharm$ set to five. Some coils exhibit strong oscillations.
The oscillation period in the AC data is linked to the period of the projection angle used in the radial
acquisition \cite{Rosenzweig_IEEETrans.Med.Imag._2020}. By removing 
this frequency and the higher-order harmonics, these oscillations can be largely eliminated. 
The filtered DC component is then used for self-gating.

The background of Figure~5 (A) represents the temporal evolution of a line profile extracted from 
a real-time image reconstruction \cite{Uecker_Magn.Reson.Med._2010} of the free-running IR radial 
FLASH with 12 inversions. The line was placed in the vertical direction of the real-time image 
series where the diaphragmatic motion can be observed, as demonstrated by the white vertical line in Figure~5 (B). 
On top of the line profiles, the 
estimated respiratory signal and the signal provided by the respiratory belt are plotted. 
All motion signals have been scaled for better visual comparison. The estimated respiratory
signal coincides well with the motion of the diaphragm in the real-time images. The adapted 
SSA-FARY technique could also provide reliable motion signal in the region where the respiratory belt failed
to produce a signal (pointed out by a white arrow). Figure~4 (B) shows 
the corresponding steady-state images reconstructed with the non-uniform fast Fourier transform 
(nuFFT) after binning the data into 6 respiratory motion states using the estimated respiration 
signal. As indicated by the dashed lines, the different inspiration and expiration phases are well resolved. 
Figure~5 (C) and (D) show a similar comparison for the pig experiment where reliable respiratory signal can 
be obtained, suggesting robustness of the adapted SSA-FARY technique.

\subsubsection*{Model-Based Myocardial $T_{1}$ Mapping}

We validated the effects of different regularization types used in the model-based reconstruction. Figure~6 shows
myocardial $T_{1}$ maps (at the selected end-expiration and end-diastolic phase) for one subject and the corresponding 
$T_{1}$ line profiles (as indicated by the dashed black line) through all cardiac phases with various regularization types. 
The regularization parameter $\alpha_{\min}$ has been optimized for each type of reconstruction individually. 
In particular, $\alpha_{\min}$ was set to be 0.02, 0.02, 0.006 and 0.005 for spatial ($\ell_{1}$-Wavelet) only, 
temporal TV only, combined spatial ($\ell_{1}$-Wavelet) and temporal TV, and the proposed regularization that combines 
spatial ($\ell_{1}$-Wavelet) and a spatio-temporal TV, respectively. The spatial ($\ell_{1}$-Wavelet) only 
regularization creates noisy and degraded myocardial $T_{1}$ maps. The temporal TV regularization could improve the 
image quality significantly by exploiting temporal sparsity. However, this kind of regularization also favors 
straight lines and thus creates "line"-like artifacts along the time/motion dimension (as seen the line profile images). 
The combination of spatial ($\ell_{1}$-Wavelet) and temporal TV regularization could reduce these "line"-like effects as 
a weaker temporal TV regularization is sufficient to achieve a similar denoising effect. Finally, the
spatio-temporal TV regularization that combined both spatial and temporal information in a single
multi-dimensional TV regularization, and its combination with the 
spatial ($\ell_{1}$-Wavelet) regularization, could achieve an even better compromise between denoising 
and the preservation of subtle motion in the line profiles (indicated by the black arrows) than the other types of regularization. 
Noteworthy, the myocardial $T_{1}$ map from the spatial-only 
regularization is noisier than our previous single-shot results \cite{Wang_J.Cardiovasc.Magn.Reson._2019}. This is 
mainly due to the fact that there is much less data in one motion state in the motion-resolved reconstruction than 
the one in \cite{Wang_J.Cardiovasc.Magn.Reson._2019} which combines data from several diastolic phases (e.g., $\sim$285 
spokes vs $\sim$800 spokes). The Supporting Information Figure~S2 shows a similar comparison of the effects of 
various regularization types on a simulated dynamic phantom with three small tubes on the "myocardium", mimicking certain "lesions". 
Here, all reconstructions were done with the same 
regularization parameter. In line with the in vivo results presented here, the spatial regularization-only reconstruction 
results in blurred images with artifacts and signal inhomogeneities on the "myocardium" due to high 
undersampling. Temporal TV regularization is able to largely remove the above artifacts and improve image 
sharpness by exploiting the temporal sparsity but favors "line"-like artifacts along the motion dimension. 
On the contrary, the proposed spatio-temporal TV combined with the 
spatial ($\ell_{1}$-Wavelet) regularization has the best performance in denoising and preservation 
of both spatial and temporal structure details.

Figure~7(A) shows the effects of the minimum regularization parameter $\alpha_{\rm min}$ on myocardial
$T_{1}$ maps and the line profiles through the cardiac phases using the combination of the
spatio-temporal TV and the spatial ($\ell_{1}$-Wavelet) regularization. Figure~7 (B) 
presents the corresponding quantitative myocardial septal $T_{1}$ values for the ROI.
As expected, both qualitative and quantitative results reveal that low values of $\alpha_{\rm min}$ result
in noisy maps (higher standard deviation) while high values may introduce blurring in the images.
$\alpha_{\rm min} = 0.005$ was then chosen to balance noise reduction and preservation of anatomical details.

With the above settings, Figure~8 shows a MOLLI $T_{1}$ map and two mid-ventricular myocardial $T_{1}$ maps 
at the end-diastolic and end-systolic phases (the same respiratory motion state) as well as the $T_{1}$ 
line profile through the cardiac phase using the proposed method for two representative subjects.  
Although the breathing conditions are different, diastolic myocardial $T_{1}$ maps are visually comparable between MOLLI
and the free-breathing technique. Besides the diastolic $T_{1}$ map, the proposed method could also provide myocardial $T_{1}$ maps 
at other cardiac phases. 

The Supporting Information Figure~S3 then presents two repetitive mid-ventricular myocardial $T_{1}$ 
maps (end-expiration and end-diastolic) of the proposed method and a MOLLI $T_{1}$ map for all subjects. 
Despite differences in breathing conditions between scans, the free-breathing $T_{1}$ maps are visually 
comparable between the two repetitive scans for all subjects. Figure~9 (A) shows the Bullseye plots of 
quantitative $T_{1}$ values and measurement repeatability errors for the six mid-ventricular 
segments of all subjects and scans for both the proposed motion-resolved model-based reconstruction and MOLLI techniques. 
Figure~9 (B) compares diastolic $T_{1}$ values for all mid-ventricular segments and septal segments
(segments 8 and 9 according to AHA) for both methods. The paired t-test comparisons for each 
segment are summarized in the Supporting Information Table~S3. 
The above quantitative comparison demonstrates that
the proposed technique has slightly shorter mean $T_{1}$ values for all 
segments ($1218 \pm 56$ ms vs $1231 \pm 40$ ms) but longer $T_{1}$ values for the septum 
segments ($1262 \pm 38$ ms vs $1250 \pm 28$ ms) than MOLLI. However, no significant differences
were found in most of the AHA segments, except for the lateral segments
where MOLLI has longer $T_{1}$ values. Noteworthy, all the above $T_{1}$ values are within 
the published normal range at 3T \cite{von_JCMR_2013}. Moreover, the proposed method has 
a slightly lower $T_{1}$ precision (higher CoV values) than MOLLI (CoV: $4.5\% \pm 1.4\%$ vs $2.8\% \pm 1.1\%$, $p < 0.01$) 
but are comparable to MOLLI in the 
repeatability errors ($34 \pm 12$ ms vs $31 \pm 13$ ms, $p = 0.73$) for all mid-ventricular 
segments. The Bland-Altman plot in the Supporting Information 
Figure~S4 further reveal that the proposed $T_{1}$ correction formula generates longer $T_{1}$ values than 
the conventional Look-Locker correction technique ($1262~\pm~38$ ms vs $1238~\pm~35$ ms, $p$ < 0.01). 

Representative basal and apical diastolic myocardial $T_{1}$ maps, in addition to the 
mid-slice $T_{1}$ map,  from two subjects, are shown in the Supporting Information 
Figure~S5 (A). Quantitative 
results from both basal and apical slices and their comparison to MOLLI are presented in the Supporting Information 
Figure~S5 (B) and (C). Again, although slight mean $T_{1}$ difference 
is observed between the motion-resolved model-based reconstruction and MOLLI techniques, no significant differences 
were found in all basal and apical AHA segments as shown in the Supporting Information Table~S3. The Supporting Information Table~S4 
further shows the relative difference of the left-ventricular area calculated from the myocardial $T_{1}$ maps. 
We observe good repeatability (repeatability error: 3\%) between scans. Although the results are obtained from
a single slice, they are generally in the expected range for LVEF values.
In addition, the quantitative $T_{1}$ maps and ROI-analyzed $T_{1}$ values in the Supporting Information Figure~S6
demonstrate good agreement between the proposed approach and MOLLI for the pig experiment, i.e., both methods 
show higher and similar myocardial $T_{1}$ values in the infarcted septal and anterior wall regions,
suggesting robustness of the proposed approach.

Aside from quantitative myocardial $T_{1}$ maps, Figure~10 presents synthesized $T_{1}$-weighted cardiac 
images (bright blood and dark blood) at the two cardiac phases for the same subjects shown in Figure~8. 
Both the bright-blood and dark-blood-weighted images clearly resolve the contrast between myocardium and blood pool.
The synthetic images and myocardial $T_{1}$ maps at all cardiac phases were then converted into movies
which are available as Supporting Information Videos~S1 and~S2. Similarly, a synthetic image series for all inversion
times and all cardiac  phases of one subject can be found in the Supporting Information Video~S3.

\section{Discussion}
	In this work, we have developed a free-breathing high-resolution myocardial 
$T_{1}$ mapping technique using a free-running inversion-recovery
radial FLASH sequence and a calibrationless motion-resolved model-based reconstruction. Instead of continuous 
acquisitions, we adopt a delay time between inversions to encode $T_{1}$ information and have derived 
a correction procedure for accurate $T_{1}$ estimation without needing full $T_{1}$ recovery 
or additional $B_{1}^{+}$ mapping. We further adapted the SSA-FARY strategy 
for robust respiratory motion signal estimation from the zero-padded AC region where the trajectory-dependent 
oscillations and contrast changing signal (due to inversion) have been eliminated in preprocessing. 
Following self-gating and data sorting, we propose to estimate both parameter maps and coil sensitivity maps 
of the desired motion states directly from k-space using an extended motion-resolved model-based 
reconstruction. The latter avoids any coil calibration and can employ high-dimensional spatio-temporal 
TV regularization, in addition to the spatial regularization, to improve 
precision in $T_{1}$ while preserving the spatio-temporal information. Studies have been performed on an 
experimental phantom, eleven healthy subjects and one young landrace pig with infarcted myocardium. 

The phantom results demonstrate good $T_{1}$ accuracy of the proposed approach
over a wide range of $T_{1}$ times. 
In vivo studies have shown similar diastolic myocardial $T_{1}$ values between the proposed approach 
and MOLLI for all segments, 
except for the lateral segments. The $T_{1}$ difference in the lateral regions between the two approaches and
the difference between lateral 
and septal $T_{1}$ values can also be seen in other $T_{1}$ mapping techniques using continuous acquisitions, such as 
MR multitasking \cite{Christodoulou_NatureBiomedicalEngineering_2018} and ref.\cite{Becker_Magn.Reson.Med._2019}. 
The origin of these differences might be through-plane myocardial motion, which makes the lateral segments 
violate the assumed signal model. If new spins that experienced $T_1$ instead of $T_1^{*}$ relaxation
move into the imaging plane due to
the through-plane motion, the total signal intensity will increase, resulting in a faster signal recovery, i.e., a shorter apparent $T_{1}$ time.
This is similar to the in-flow effects in blood $T_{1}$ estimation, as analyzed by Hermann et al. \cite{Hermann_Phys.Med.Biol._2020}. Although $T_{1}$ values 
in this work correspond well with MOLLI, the proposed approach may still underestimate $T_{1}$ when compared to the saturation recovery-based approaches such as SASHA and SAPPHIRE. With a low flip-angle FLASH readout, the proposed sequence should be robust to $B_{1}$ and slice profile effects \cite{Tran-Gia_MRI_2014}. The main contributing factor for the underestimation could be imperfect inversion caused by the non-selective hyperbolic secant pulse we used. The lateral regions are additionally affected by through-plane motion as explained above. The precision of the proposed method (CoV: $4.5\% \pm 1.4\%$) is lower than that of MOLLI (CoV: $2.8\% \pm 1.1\%$). 
Such a difference could be explained by the differences in the nominal spatial resolution (MOLLI: $1.4 \times 1.4 \times 8$ $\text{mm}^{3}$, 
the proposed: $1.0 \times 1.0 \times 6.0$ $\text{mm}^{3}$) and the readouts (MOLLI: Cartesian balanced SSFP, 
the proposed: radial FLASH). Nevertheless, the proposed approach shows comparable (CoV: $4.9\%$ with 
$1.3 \times 1.3 \times 8$ mm$^{3}$ in ref. \cite{Becker_Magn.Reson.Med._2019}, CoV: $4.8\%$ 
with $1.7 \times 1.7 \times 8$ mm$^{3}$ in MR multitasking \cite{Christodoulou_NatureBiomedicalEngineering_2018}) 
or even slightly better (CoV: $5.7\%$ with $1.6 \times 1.6 \times 8$ mm$^{3}$ in 
MRF \cite{Hamilton_Magn.Reson.Med._2017}) $T_{1}$ precision when comparing to other well-known techniques.

Continuous acquisitions with constant flip angles \cite{Shaw_Magn.Reson.Med._2019, 
	Qi_Magn.Reson.Med._2019} have been used in several inversion-prepared free-running 
$T_{1}$ mapping techniques attributed to the scan efficiency. I.e., there is no waiting 
time between inversions.
However, as pointed out by \cite{Zhou_Magn.Reson.Med._2021}, continuous acquisition
with the same flip angle
only encodes $T_{1}^{*}$ information in the data \cite{Look_Rev.Sci.Instrum._1970}. Since $T_{1}^{*}$ 
is a function of flip angle and $T_{1}$, 
additional information about $B_{1}^{+}$ is therefore necessary for accurate $T_{1}$ estimation.
However, the additional estimation of $B_{1}^{+}$ at the same motion state might be difficult
to achieve in free-breathing, self-gated acquisitions. 
In \cite{Zhou_Magn.Reson.Med._2021}, Zhou et al. \cite{Zhou_Magn.Reson.Med._2021} propose to solve this issue by introducing 
a dual flip-angle strategy which acquires data continuously with two flip angles 
consecutively applied. Following self-gated data sorting, image reconstruction and 
dictionary matching, two different $T_{1}^{*}$ maps from the same motion state can be 
extracted and subsequently be used to calculate both $T_{1}$ and $B_{1}^{+}$ maps in an iterative manner. 
Most recently, a similar idea has been proposed in the MR multitasking technique for more 
accurate $T_{1}$ mapping \cite{Serry_Magn.Reson.Med._2021}.
%
Alternatively, in this work, we propose to resolve this problem by adopting a delay 
time between inversions, using this period for encoding $T_{1}$ information in the data. 
Different from studies which set the delay time long enough
to ensure a full recovery of longitudinal magnetization, we are capable of 
estimating accurate $T_{1}$ even with incomplete $T_{1}$ recovery, shortening the acquisition time. 
Moreover, the proposed approach requires neither 
additional $B_{1}^{+}$ mapping nor the explicit calculation of $B_{1}^{+}$ from the data. 

Self-gating constitutes another key component for free-breathing imaging. Although a few self-gating techniques 
have been successfully developed for steady-state imaging, estimation of reliable motion signals from the 
contrast-modulated k-space is challenging. In this work, following removal of signal oscillations in 
the k-space center signal, we model the additive and multiplicative effects caused by inversion in the data 
following \cite{Winter_Magn.Reson.Med._2016} and propose to reduce such effects prior to the application 
of the SSA-FARY-based self-gating techniques. From our experience, the above step is crucial for reliable 
motion estimation using SSA-FARY. Although the proposed method could achieve 
robust respiration signal estimation, determination of reliable cardiac signals from the filtered k-space remained
challenging and retrospective ECG gating was used for binning.
Resolving the latter issue in future work would be valuable as     
the ECG signal is not always reliable \cite{Larson_Magn.Reson.Med._2004}, which we also observed for 
several data sets acquired for this study.

Inspired by the high-dimensional imaging techniques  
\cite{Feng_Magn.Reson.Med._2016a, Cheng_J.Magn.Reson.Imaging_2016, Cheng_Sci.Rep._2017, Christodoulou_NatureBiomedicalEngineering_2018, Di_Sopra_Magn.Reson.Med._2019},
we have sorted the data into multiple
cardiac and respiratory motion states, and applied high-dimensional regularization 
along these motion dimensions to improve $T_{1}$ accuracy and precision. In contrast, 
several other studies \cite{Qi_Magn.Reson.Med._2019, Zhou_Magn.Reson.Med._2021}
combined data from multiple respiratory motion states into 
one using rigid image registration, following respiratory motion field being estimated from low-resolution images. 
The latter strategy has the advantage that more data is available for each cardiac phase than the one that sorts 
the data into multiple respiratory and cardiac
motion states within the same amount of time. 
However, as motion between respiratory states is usually considered to be
nonlinear \cite{Hansen_Magn.Reson.Med.2012} for cardiac imaging, a linear model may cause data mismatch in the cost 
function, resulting in reconstruction errors. Most recently, advanced nonlinear motion estimation methods have been
developed for whole-heart coronary MR imaging \cite{Qi_IEEETrans.Med.Imag._2020}. Integration of such a nonlinear 
motion model into the model-based reconstruction framework would also be of great interest as it has the potential 
to shorten the total acquisition time of the proposed method while preserving good $T_{1}$ accuracy and precision. 

Spatio-Temporal regularization has been shown to be more effective in exploiting sparsity in compressed-sensing reconstructions for  
dynamic/high-dimensional imaging, resulting in higher accelerator factors than spatial regularization-only reconstruction \cite{Feng_Magn.Reson.Med._2016a, Cheng_J.Magn.Reson.Imaging_2016,Christodoulou_NatureBiomedicalEngineering_2018}. 
This work confirms the above findings in the regularized nonlinear model-based reconstruction for dynamic myocardial $T_{1}$ mapping. Moreover, 
our results demonstrate that the spatio-temporal TV 
regularization has a slightly better performance in both image denoising and preservation of structure details than the temporal TV regularization. 
On the other hand, although the regularization used in this study is effective in reducing noise/improving quantitative precision, 
it may also cause a certain degree of image blurring (lower effective spatial resolution) similar to other 
regularization techniques used in compressed sensing. More advanced regularization, such as neural network-enhanced 
regularizers \cite{Hammernik_Magn.Reson.Med._2018} could be employed in future studies to solve this issue. 

The proposed method takes around two minutes for reliable $T_{1}$ estimation, which compares well to alternative
techniques when considering the relatively high nominal resolution of the $T_1$ maps ($1.0 \times 1.0 \times 6$ mm$^{3}$).
The acquisition time could be shorted by further reducing the delay time. Here, we adopted the three-second delay
to achieve a good compromise of $T_{1}$ accuracy and precision. However, in principle, a delay time of one
second could be used (at a cost of lower precision), resulting in acquisition times of around 80 s.

There are also other limitations of the present work that need 
to be mentioned. First, the blood $T_{1}$ estimated by methods using continuous acquisition
may not be reliable as the in-flow effects make the blood violate the assumed signal model,
a problem which also affects other methods based on continuous acquisition
\cite{Hermann_Phys.Med.Biol._2020, Zhou_Magn.Reson.Med._2021, Qi_Magn.Reson.Med._2019}. 
Thus, the proposed method is not ideal for estimating the extracellular volume (ECV). A thorough investigation of how blood $T_{1}$ is affected by the proposed sequence using simulations and a flow phantom, similar to the work in \cite{Hermann_Phys.Med.Biol._2020}, would be an interesting next step. 
Second, evaluation of the proposed method has so far only been done 
in healthy volunteers. Validation of the proposed free-breathing method in
patient studies with both native and 
post-contrast $T_{1}$ mapping is now warranted and will be the subject of future work.
Another limitation of the proposed method is the long computation time. Although substantial efforts 
have been made in the implementation part to enable model-based reconstruction to run on GPUs,
which already reduced reconstruction time from several hours to 25 minutes, further
efforts are still needed.

\section{Conclusion}
The proposed free-breathing method enables high-resolution $T_{1}$ mapping 
with good $T_{1}$ accuracy, precision and repeatability by combing inversion-recovery
radial FLASH, self-gating and a calibrationless motion-resolved model-based reconstruction.

\section{Acknowledgement}
We thank Dr. Haikun Qi from ShanghaiTech University for insightful discussions.

\section{Funding Information}

This work was supported by the DZHK (German Centre for Cardiovascular Research), by the Deutsche Forschungsgemeinschaft (DFG, German Research Foundation) grants - UE 189/1-1, UE 189/4-1, TA 1473/2-1, and EXC 2067/1-390729940, and funded in part by NIH under grant U24EB029240. This project has also received funding from the European Union's Horizon 2020 research and innovation program under grant agreement No. 874764.

\section*{Open Research}
\subsection*{Data Availability Statement}
In the spirit of reproducible research, code to
reproduce the experiments will be available on
\url{https://github.com/mrirecon/motion-resolved-myocardial-T1-mapping}.
The raw k-space data, all ROIs to reproduce the quantitative values and other relevant files used in this study can be downloaded from
\url{https://doi.org/10.5281/zenodo.5707688} and \url{https://doi.org/10.5281/zenodo.7350323}.
\appendix
\section{Appendix}

\subsection*{IRGNM-ADMM algorithm}


\begin{algorithm*}[H]
	\caption{IRGNM-ADMM algorithm}\label{alg:irgnm-admm}
	$\textbf{OUTPUTS:}$ $(M_{ss}, M_{0}^{'}, R_{1}^{*})^{T}$ and $(c_{1},\cdots,c_{N_{\text C}})^{T}$ for all motion states\\
	$\textbf{INPUTS:}$ \\
	$Y \gets \textnormal{Gridded and sorted k-space data}$\;
	$P \gets \textnormal{Sampling pattern}$\;
	$(t_{1}, \cdots, t_{n_{\text{TI}}})^{T} \gets \textnormal{Vector of inversion times for each motion state}$\;
	
	%
	\text{\% Initialization for IRGNM}:\\
	$n = 0$, $\alpha_{0} = \beta_{0} = 1$, \textnormal{MaxIter} = 10, $x_{0} = (1, 1, 1.5, 0, \cdots, 0)^{T}$\\
	\text{\% Define $A$ to be the block diagonal matrix with the blocks $A_{r,c}$ on the diagonal and} \\ {\% $b$ the stacked vector of $b_{r,c}$ where}
	\begin{align*}
	A_{r,c} &= DF_{r,c}(x_{n}) \\
	b_{r,c} &= DF_{r,c}(x_{n})x_{n} -F_{r,c}(x_{n}) + {Y}_{r,c}
	\end{align*}
	\text{\% with $DF_{r,c}(x_{n})$ the Frech\'et derivative of $F_{r,c}$ at the point $x_{n}$ for the $n$th 
		Gauss-Newton step.}
	
	\text{\% Gauss-Newton Iterations:} \\
	\While{$n < \textnormal{MaxIter}$}{ 
		\text{\% Solve the following linearized subproblem with ADMM:}\\
		\text{\%} ${x}_{n+1} = \argmin_{x \in S} \|Ax-b \|_{2}^{2} + \alpha_{n} R(x_{\bm{m}}) + \beta_{n} U(x_{\bm{c}})$, \text{ with $x = (x_{\bm{m}}, x_{\bm{c}})^{T}$}. \\
		\text{\% Initialization for ADMM}:\\
		$k$ = 0, K = \text{min}(100, $10 \cdot 2^{n}$), $\rho = 0.01$, $z^{k} = y^{k} = x_{n}$.\\
		\text{\% ADMM Iterations:} \\
		\For{$k < \textnormal{K}$}{ 
			$x^{k+1} \gets (A^{H}A + \rho I)^{-1} (A^{H}b +\rho z^{k} - y^{k})$ \text{\% solved by conjugate gradient}\;
			$z_{\bm{m}}^{k+1} \gets \textbf{prox}^{\bm{m}}_{\alpha_{n}/\rho}(x_{\bm{m}}^{k+1} + y_{\bm{m}}^{k}/\rho)$  	\text{\% proximal operators for parameter maps} $x_{\bm{m}}$\;
			$z_{\bm{c}}^{k+1} \gets 			\textbf{prox}^{\bm{c}}_{\beta_{n}/\rho}(x_{\bm{c}}^{k+1} + y_{\bm{c}}^{k}/\rho)$ \text{\% proximal operators for coil sensitivity maps} $x_{\bm{c}}$\; 
			$ y^{k+1} \gets y^{k} + \rho(x^{k+1} - z^{k+1})$\;
		}
		$x_{n+1} = y^{k+1}$\;
		$\alpha_{n+1} = \text{max}\large(\alpha_{\text{min}}, (\frac{1}{3})^{n}\cdot \alpha_{0}\large)$\;
		$\beta_{n+1} = (\frac{1}{3})^{n}\cdot \beta_{0}$\;
		n = n + 1\;
	}
\end{algorithm*}

In the above algorithm, $\textbf{prox}^{\bm{m}}_{\alpha_{n}/\rho}$ contains the following three proximal operators:
\begin{align*}
Z_{\bm m}^{k+1} &:= W^{H}S_{\alpha_{n}/\rho}(WZ_{\bm m}^{k})\quad (\text{Wavelet-domain joint soft-thresholding}) \\
Z_{\bm m}^{k+1} &:= P_{S}(Z_{\bm m}^{k}) \quad (\text{projection of $Z_{\bm m}^{k}$ onto domain $S$}) \\
Z_{\bm m}^{k+1} &:=  D^{H}S_{\alpha_{n}/\rho}DZ_{\bm m}^{k} \quad (\text{Total variation gradient-domain soft-thresholding})
\end{align*}
and $\textbf{prox}^{\bm{c}}_{\alpha_{n}/\rho}$ is the least-square proximal operator, i.e.,
\begin{align*}
Z_{\bm c}^{k+1} &:= \frac{1}{2}Z_{\bm c}^{k}.
\end{align*}

\bibliographystyle{mrm}
\bibliography{main}
\section*{Figures}

\begin{figure}[H]
	\centering
	\includegraphics[width=1.0\textwidth]{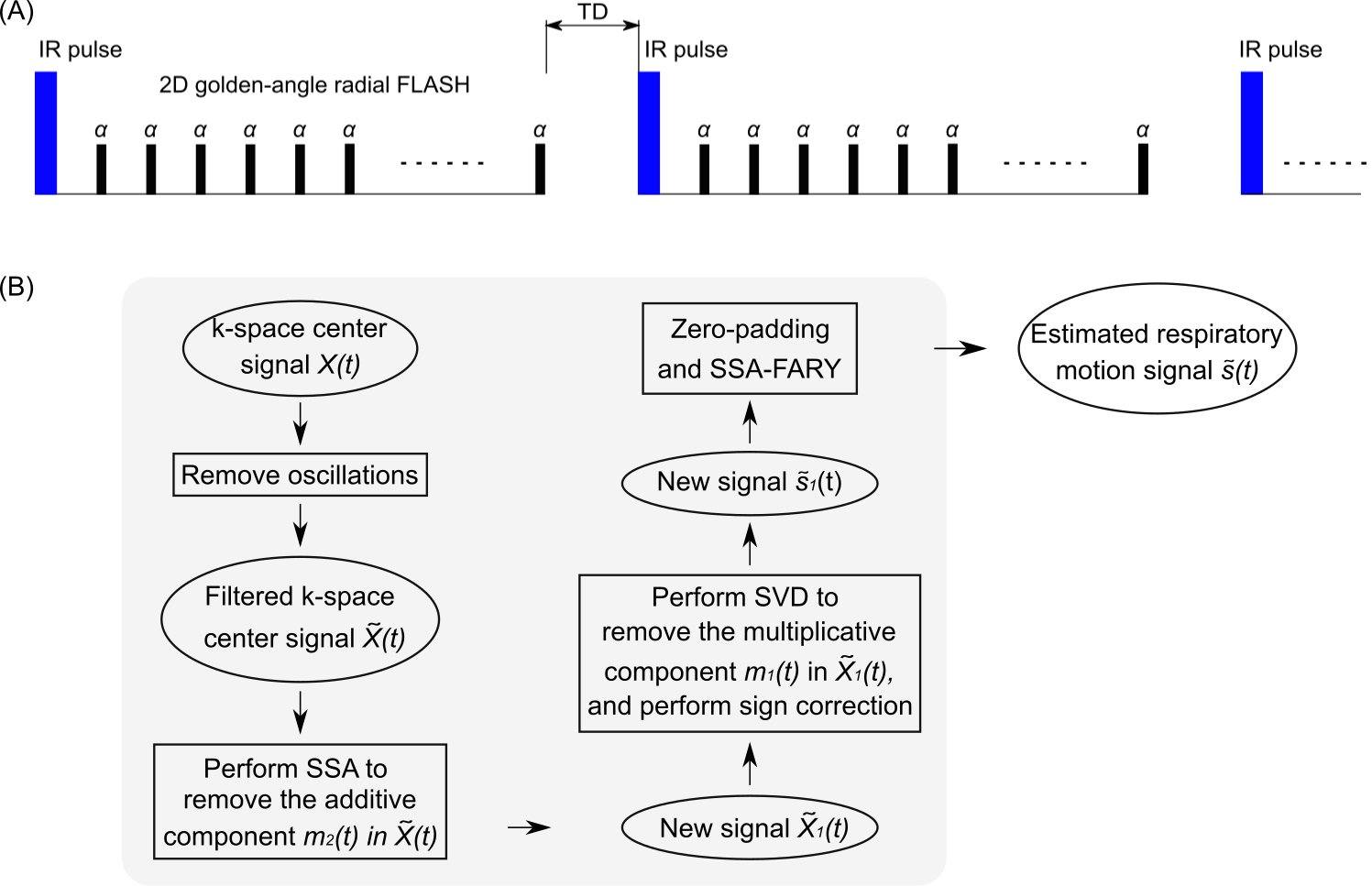}
	\caption{A. Schematic diagram of the free-running inversion-recovery radial FLASH sequence. Note TD is the delay time between inversions and this period encodes pure $T_{1}$ information in the data. B. Flowchart of the main steps in the adapted SSA-FARY technique for the respiratory motion signal estimation from the k-space center.}
	\label{scheme_fig1}
\end{figure}

\begin{figure}[H]
	\centering
	\includegraphics[width=1.0\textwidth]{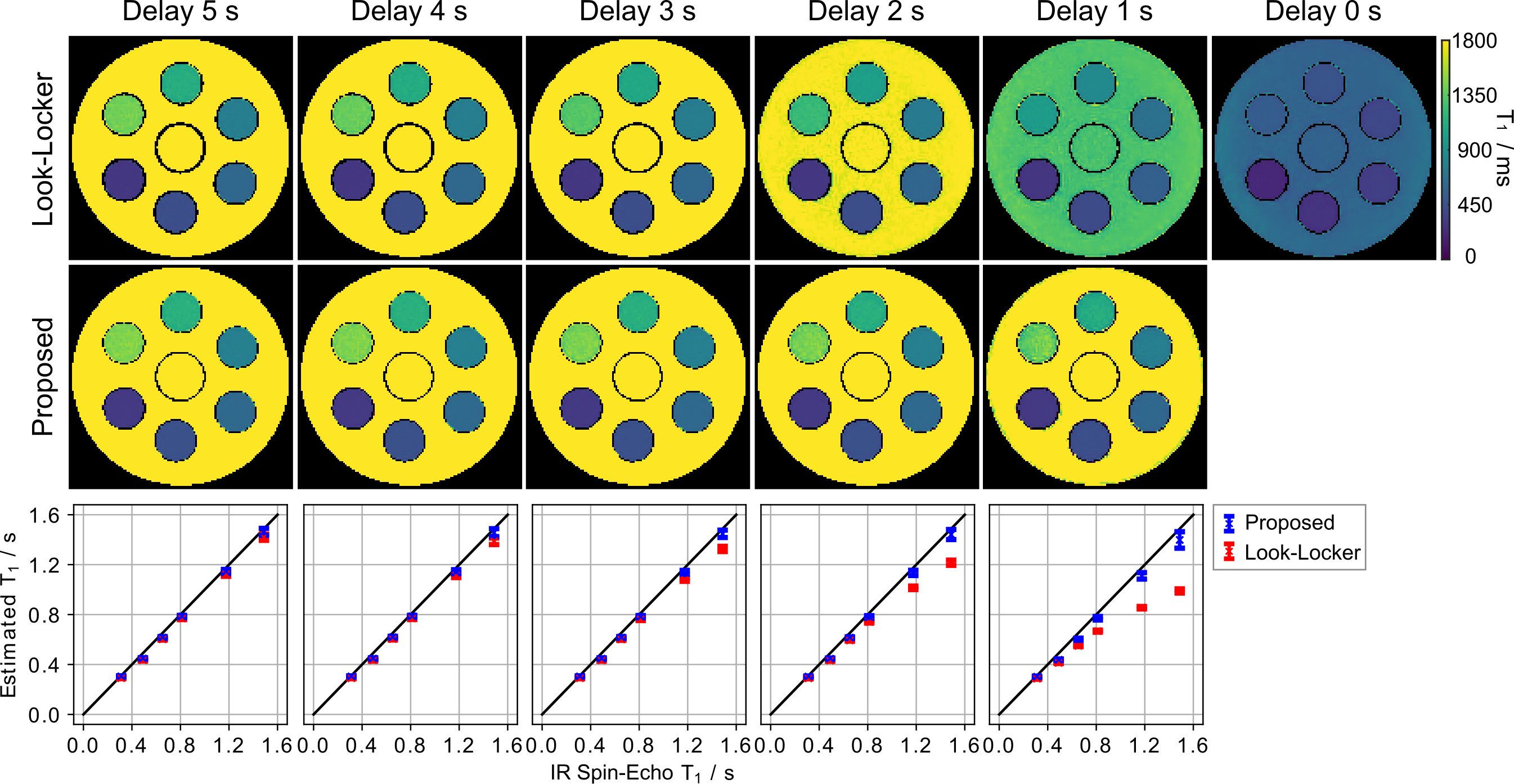}
	\caption{Quantitative phantom $T_{1}$ maps with various 
		delay times from 5 s to 1 s (step size 1 s) using (top) the conventional Look-Locker correction and (middle) the proposed 
		formula. (Bottom) Quantitative $T_{1}$ values (mean and standard deviation) within ROIs of the 6 phantom tubes in
		comparison to an IR spin-echo reference. The corresponding Bland-Altman plots are shown in the Supporting Information Figure~S1 (A).}
	\label{phantom_1}
\end{figure}

\begin{figure}[H]
	\centering
	\includegraphics[width=1.0\textwidth]{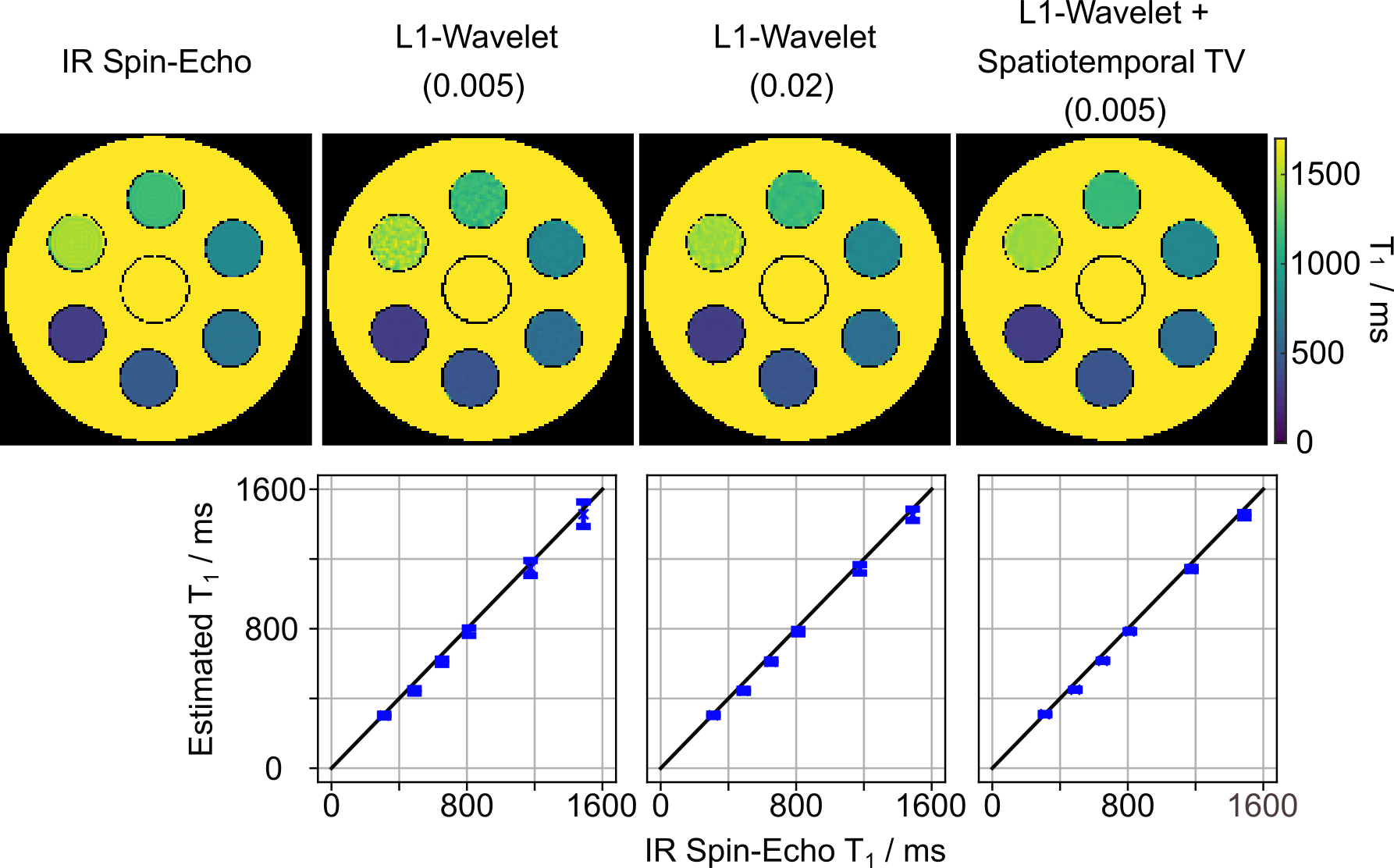}
	\caption{(Top) Phantom $T_{1}$ maps reconstructed with different regularization using the motion-resolved
		model-based reconstruction in comparison to an IR spin-echo reference. (Bottom) Quantitative $T_{1}$ values
		(mean and standard deviation) within ROIs of the 6 phantom tubes. The value in the bracket (top) indicates the
		regularization parameter $\alpha_{\min}$ used for each reconstruction. The corresponding Bland-Altman plots are presented in the Supporting Information Figure~S1 (B).}
	\label{phantom_2}
\end{figure}

\begin{figure}[H]
	\centering
	\includegraphics[width=1.0\textwidth]{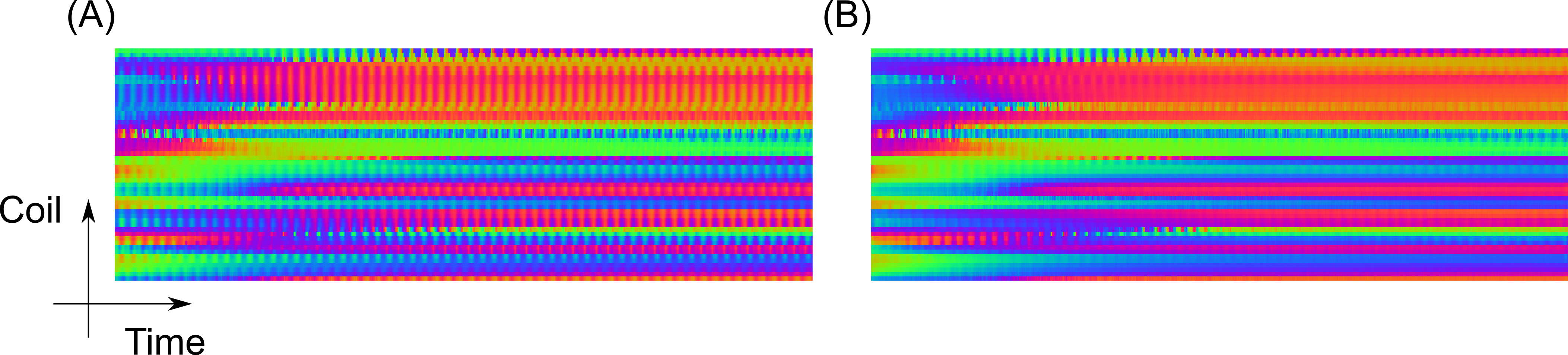}
	\caption{Snippet of the complex plot with color-coded phase of the DC samples used 
		for auto-calibration before (A) and after (B) data correction with the extended 
		orthogonal projection. Notably less disturbing oscillations are observed in B. The above snippet corresponds to one complete inversion 
		recovery (3 s). }
	\label{fig1}
\end{figure}

\begin{figure}[H]
	\centering
	\includegraphics[width=1.0\textwidth]{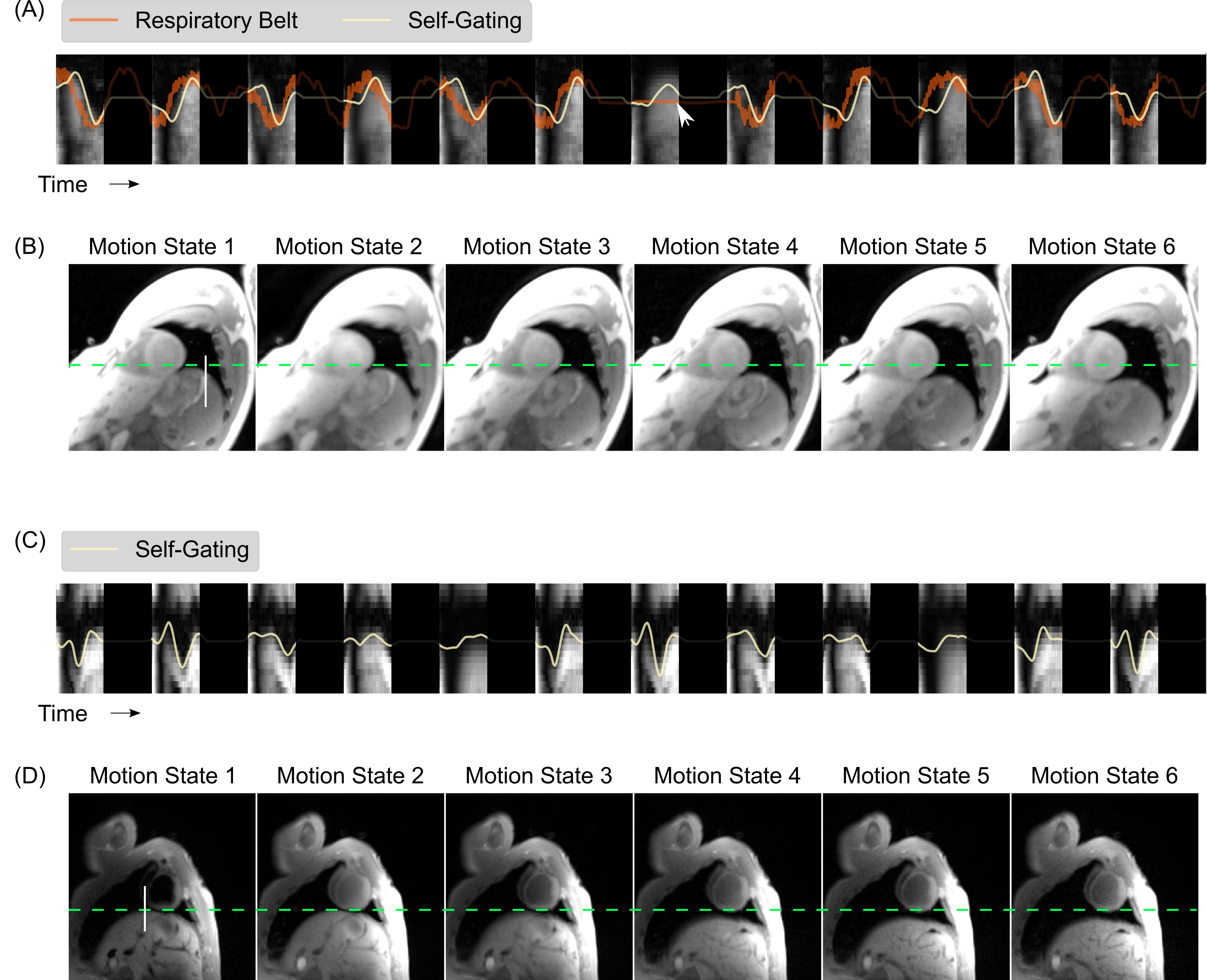}
	\caption{A. Comparison of the estimated respiratory signal with that obtained from the respiratory belt for 12 inversions for a healthy subject.
		The background image represents the temporal evolution of a vertical line profile (white line in B) extracted from 
		a real-time image reconstruction \cite{Uecker_Magn.Reson.Med._2010} of the data acquired with free-running IR radial 
		FLASH. The dark regions represent the time delay between inversions. The white arrow indicates a time point where
		the respiration belt failed to provide a signal. B. The corresponding steady-state images reconstructed by the non-uniform fast Fourier transform 
		after binning the data (combing all cardiac phases) into 6 respiratory motion states. The dashed green line serves as 
		a baseline for the end-respiration motion state. 
		(C) and (D) show similar results for the pig experiment but with the respiratory belt signal absent.}
	\label{Resp_motion_Est}
\end{figure}

\begin{figure}[H]
	\centering
	\includegraphics[width=1.0\textwidth]{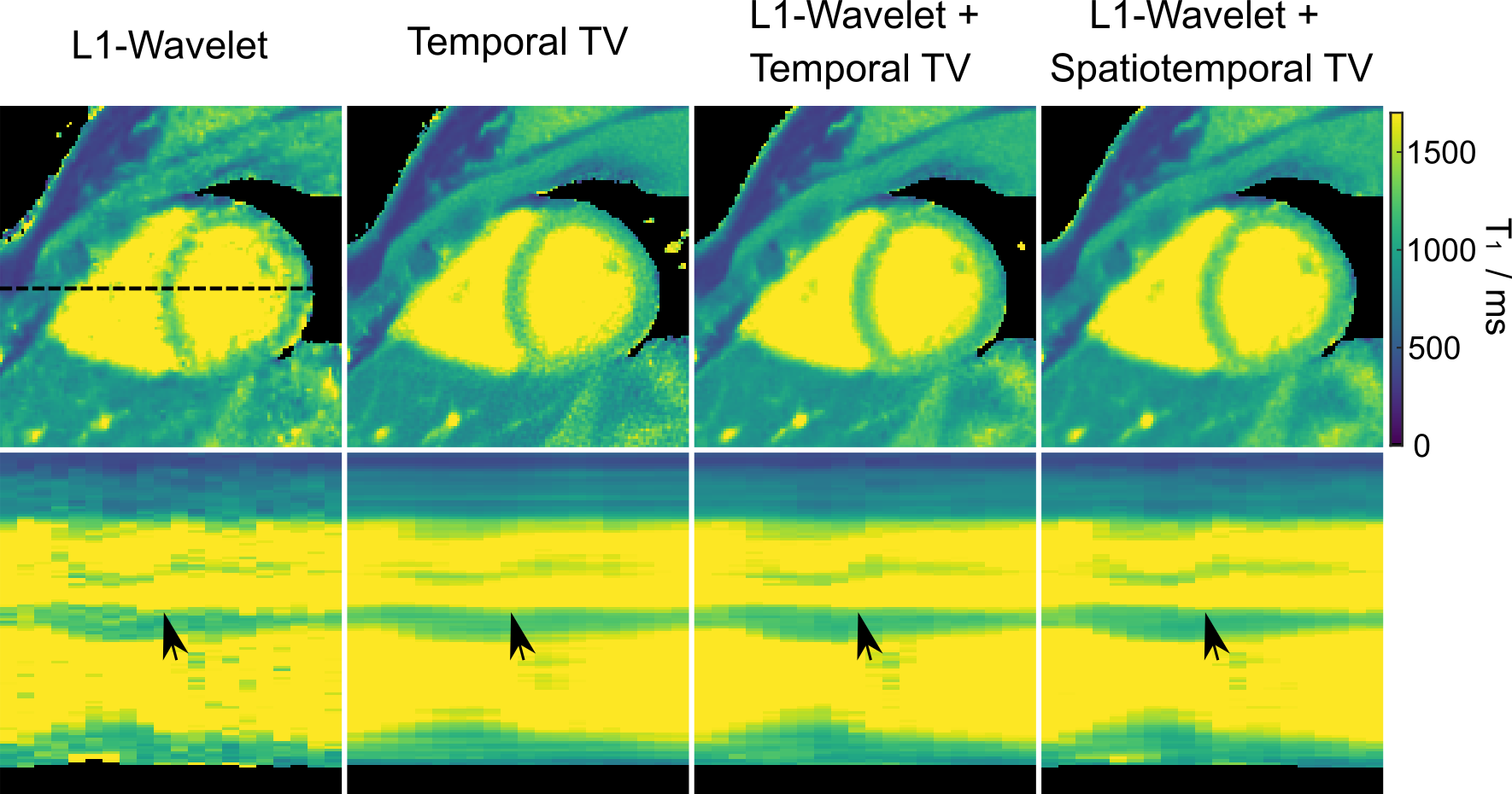}
	\caption{(Top) Myocardial $T_{1}$ maps (end-expiration and end-diastolic) with different types of regularization using the proposed 
		motion-resolved model-based reconstruction. (Bottom) Horizontal profiles (dashed black line in the top) through all cardiac phases.
		The black arrows indicate subtle wall motion that is preserved best with the spatiotemporal TV 
		regularization. Note that the regularization parameter $\alpha_{\min}$ for each regularization type was
		tuned individually to achieve a fair comparison.}
	\label{invivo 1}
\end{figure}

\begin{figure}[H]
	\centering
	\includegraphics[width=1.0\textwidth]{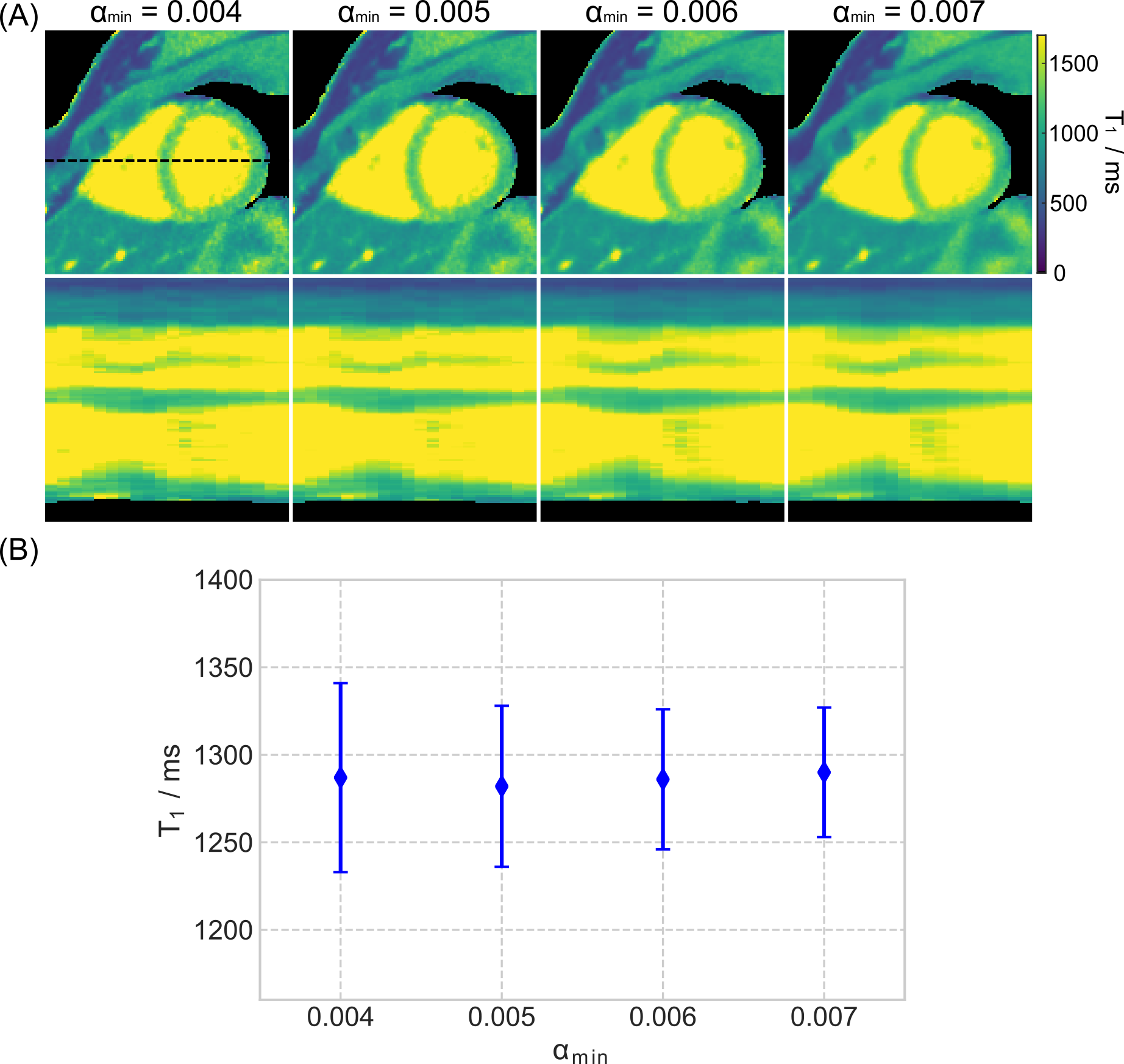}
	\caption{A. (Top) Myocardial $T_{1}$ maps (end-expiration and end-diastolic) estimated with
		motion-resolved model-based reconstruction with different choices of
		the minimum regularization parameter $\alpha_{\rm min}$. 
		(Bottom) Horizontal profiles (dashed black line in the top) through all cardiac phases.
		B. Quantitative $T_{1}$ values (mean and standard deviation) within a ROI in the septal region.}
	\label{invivo 2}
	
\end{figure}

\begin{figure}[H]
	\centering
	\includegraphics[width=1.0\textwidth]{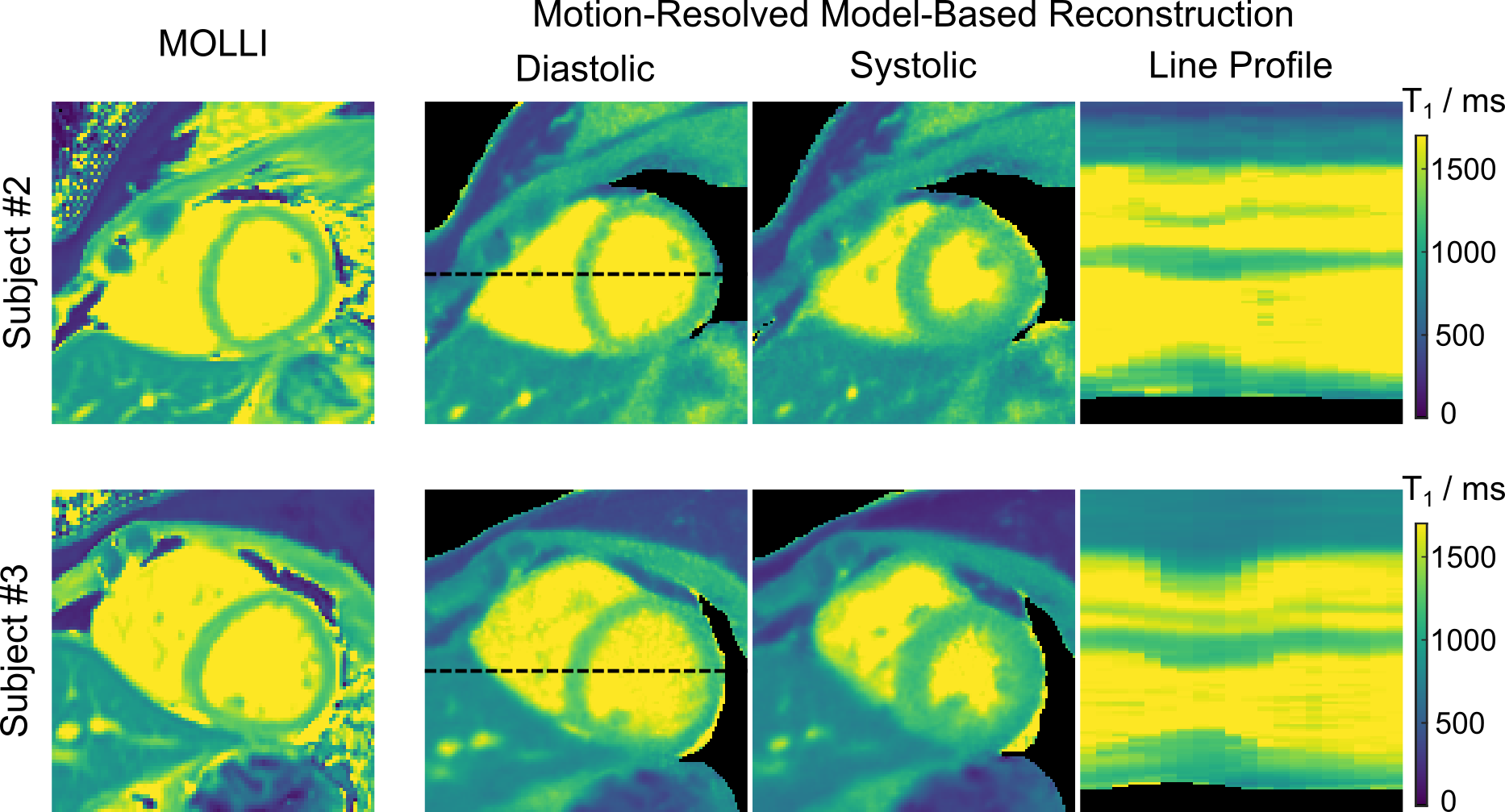}
	\caption{Diastolic and systolic myocardial $T_{1}$ maps (end-expiration) and line profiles through the cardiac phase of the
		motion-resolved model-based reconstruction acquired during free breathing
		in comparison to MOLLI acquired in a breathhold for two representative subjects. }
	\label{diast_syst_synt}
\end{figure}

\begin{figure}[H]
	\centering
	\includegraphics[width=1.0\textwidth]{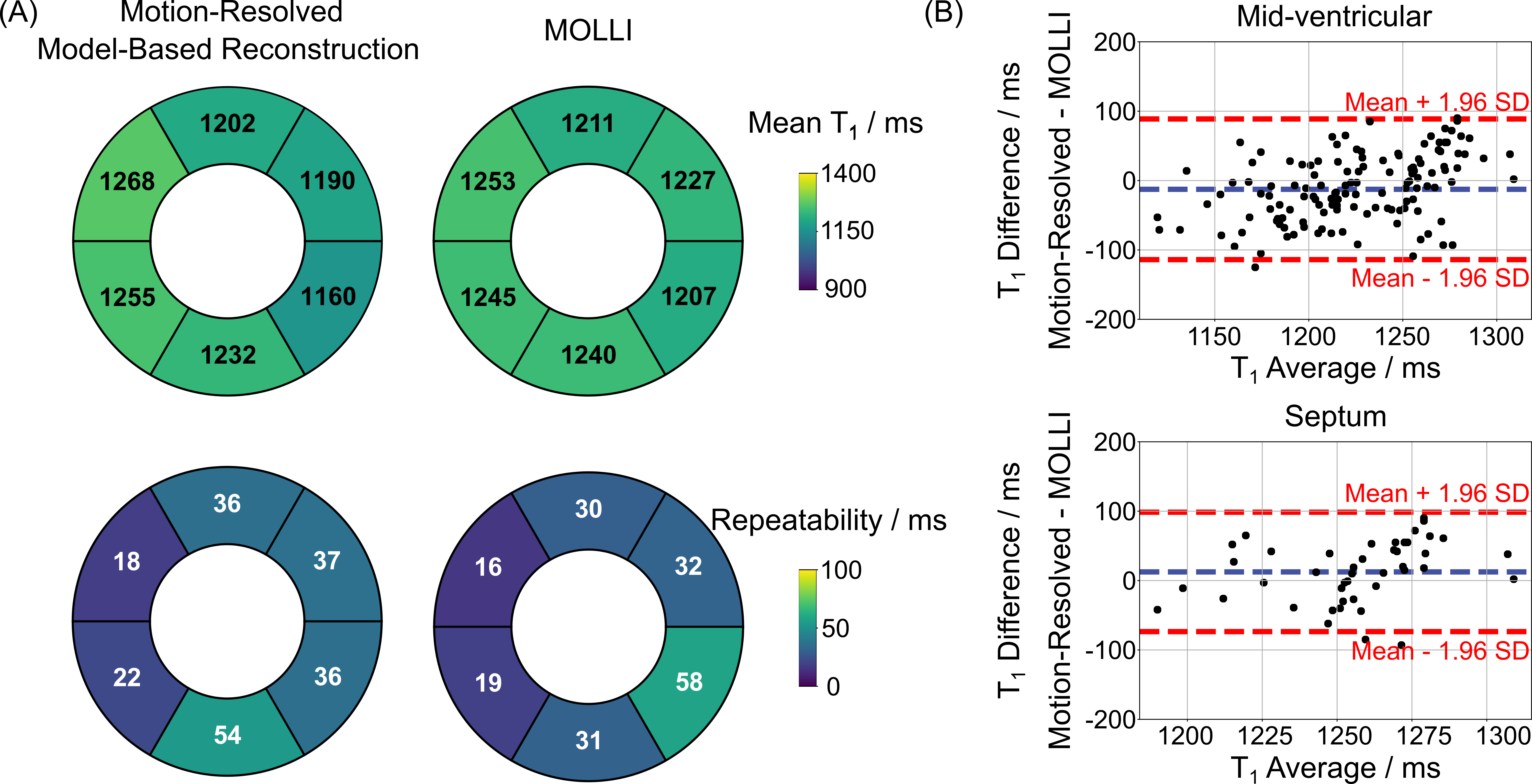}
	\caption{A. Bullseye plots of six mid-ventricular myocardial segments, showing (top) the mean diastolic $T_{1}$ 
		values and (bottom) the measurement repeatability errors for all eleven subjects and all scans for (left) 
		the motion-resolved model-based reconstruction (free-breathing)  and (right) the MOLLI method (breathhold), respectively. B. Bland–Altman plots comparing
		the mean diastolic $T_{1}$ values of (top) all the six mid-ventricular segments (mean difference: -12 ms, SD: 52 ms) and (bottom) the septal 
		segments (segments 8 and 9 according to AHA, mean difference: 12 ms, SD: 44 ms) for the proposed method and MOLLI for all subjects and scans.}
	\label{invivo_quantitative}
\end{figure}

\begin{figure}[H]
	\centering
	\includegraphics[width=1.0\textwidth]{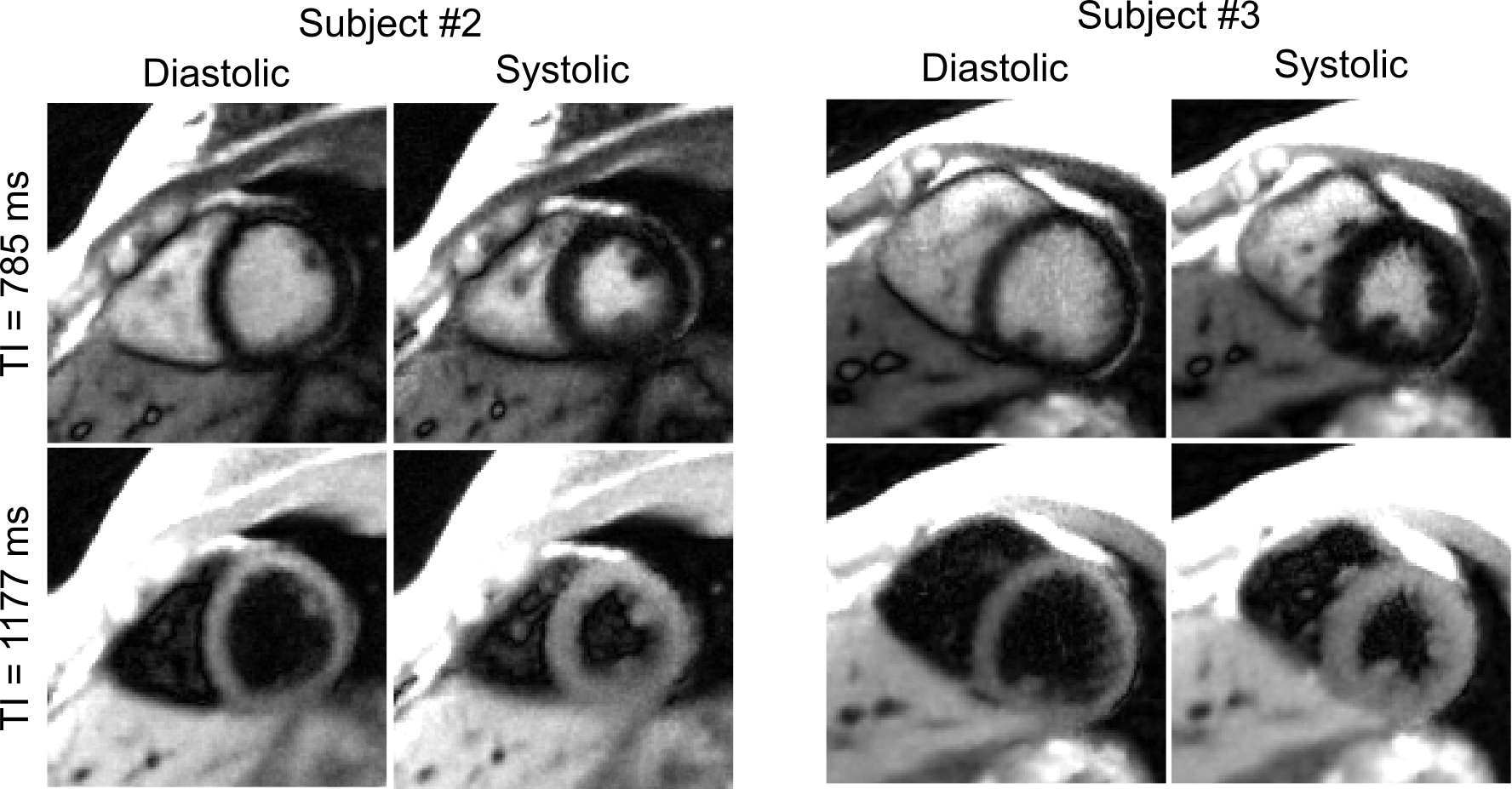}
	\caption{Synthesized $T_{1}$-weighted images at two representative inversion times (bright blood 
		and dark blood) for the end-diastolic and end-systolic cardiac phases for the same subjects shown in Figure 8.  }
	\label{invivo 4}
\end{figure}



\section*{Supplementary material}

# Free-Breathing Myocardial $T_1$ Mapping using Inversion-Recovery Radial FLASH and Motion-Resolved Model-Based Reconstruction

Wang X, Rosenzweig S, Roeloffs V, Blumenthal M, Scholand N, Tan Z,

Holme HCM, Unterberg-Buchwald C, Hinkel R, and Uecker M.

## I  Derivation of $T_1$ Estimation Formula From Incomplete (Partial) Recovery

To describe the effect of partial recovery, we split the recovery curve into two parts: The imaging part (subject to $R_1^*$ relaxation) with a time period of $t_{1s}$ and the free recovery part (subject to $R_1$ relaxation) with a period of $t_1$. The dynamics of each part is then governed by:

$$\text{Imaging}: \quad M_{ss} - (M_{ss} - M_{\text{ini}}) \cdot e^{-R_1^* \cdot t_{1s}} =: A_1 + B_1 M_{\text{ini}}$$
$$\text{Free relaxation}: \quad M_0 - (M_0 - M_{\text{ini}}) \cdot e^{-R_1 \cdot t_1} =: A_2 + B_2 M_{\text{ini}}$$

where $t_{1s}$ and $t_1$ are the time periods for imaging and free relaxation parts, respectively.

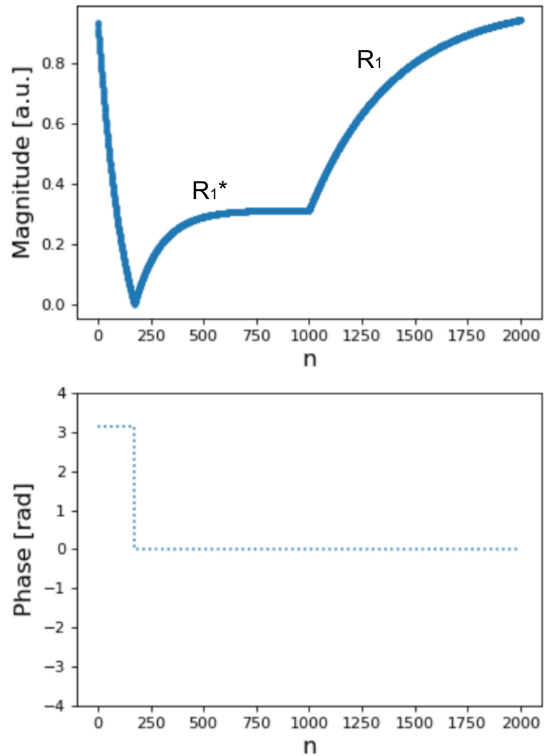

Figure 1: Demonstration of IR Look-Locker signal with partial recovery: 3 s imaging followed by 3 s free recovery. n describes the index of the RF excitation.

The joint effects of these two blocks can then be written as:

free relaxation after imaging: $A_2 + B_2(A_1 + B_1 M_{\text{ini}}) =: A_3 + B_3 M_{\text{ini}}$

where $A_3 = A_2 + B_2 A_1$ and $B_3 = B_2 B_1$

For the new signal model, we simply have to find the fix point which gets mapped to itself by both relaxation parts and the inversion:

$$M_0^{'} = A_3 + B_3 \cdot (-M_0^{'})$$
$$\Rightarrow M_0^{'} = \frac{A_3}{1 + B_3}$$
$$= \frac{A_2 + B_2 A_1}{1 + B_2 B_1}$$
$$= \frac{M_0(1 - e^{-R_1 \cdot t_1}) + e^{-R_1 \cdot t_1} M_{ss}(1 - e^{-R_1^* \cdot t_{1s}})}{1 + e^{-R_1 \cdot t_1} \cdot e^{-R_1^* \cdot t_{1s}}}$$

By substituting $M_0 = R_1^{-1} M_{ss} R_1^*$, we obtain the formula of equation (1) in the main manuscript.

Therefore, the key idea is that even with partial recovery, the signal is still mono-exponential and can be fitted with a 3-parameter model

$$s(t) = M_{ss} - (M_{ss} + M_0^{'}) \cdot e^{-t \cdot R_1^*}$$

Following parameter estimation, accurate $T_1$ can be achieved using equation (1) in the main manuscript in a post-processing step.

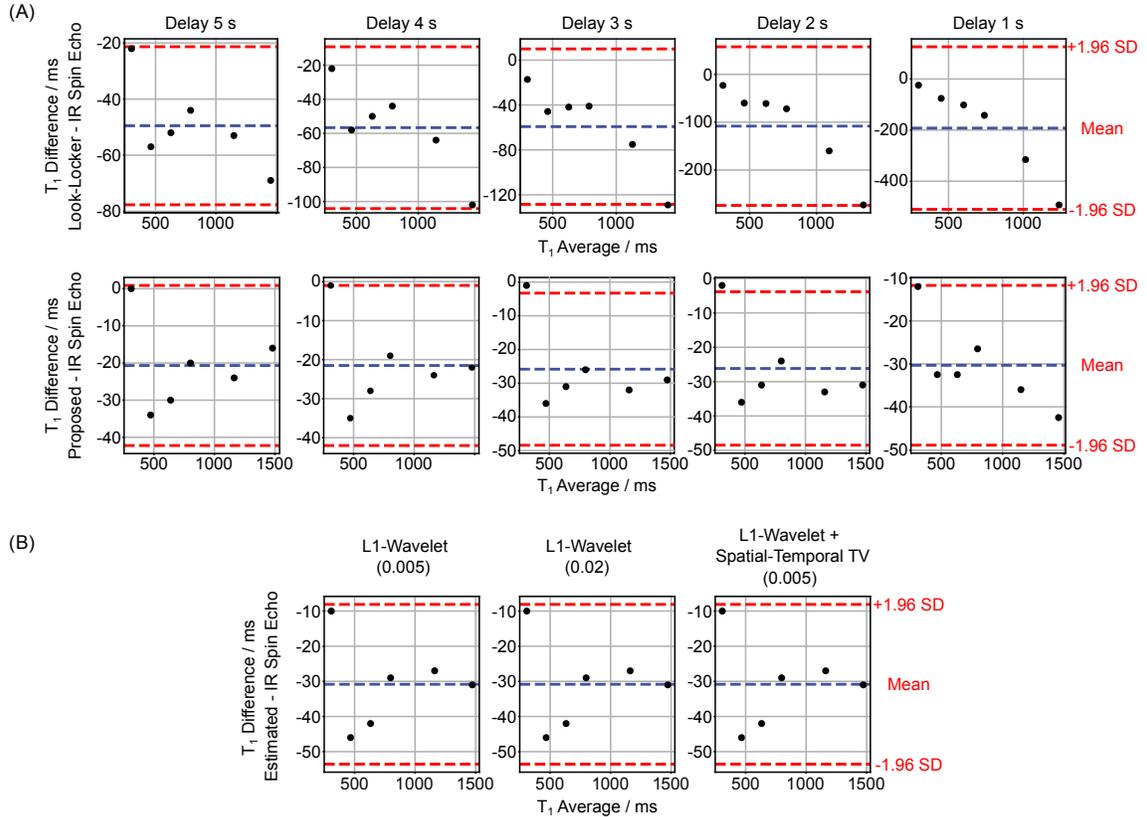

**Supporting Information Figure S1.** A. Bland-Altman plots comparing ROI-analyzed phantom $T_1$ values (top) between the original Look-Locker correction and the IR spin-echo reference (from left to right, mean difference: -50, -57, -74, -108 and -195 ms, and SD: 14, 24, 45, 85 and 163 ms), and (bottom) between the proposed correction and the IR spin-echo reference (from left to right, mean difference: -21, -22, -26, -26 and -41 ms, and SD: 11, 10, 12, 11 and 19 ms) for various delays. B. Bland-Altman plots comparing ROI-analyzed phantom $T_1$ values between motion-resolved model-based reconstruction with different regularizations and the IR spin-echo reference (from left to right, mean difference: -31, -31 and -31 ms, and SD: 12, 12 and 12 ms).

**Supporting Information Table S1.** $T_1$ relaxation times (ms, mean ± SD) and CoV values (%) for the experimental phantom in Figure 2.

|  |  | Tube 1 | | Tube 2 | | Tube 3 | | Tube 4 | | Tube 5 | | Tube 6 | |
|---|---|---|---|---|---|---|---|---|---|---|---|---|---|
|  |  | $T_1$ | CoV | $T_1$ | CoV | $T_1$ | CoV | $T_1$ | CoV | $T_1$ | CoV | $T_1$ | CoV |
| Delay 5 s | Look-Locker | 290 ± 4 | 1.4 % | 433 ± 5 | 1.2 % | 601 ± 5 | 0.8 % | 768 ± 7 | 0.9 % | 1122 ± 15 | 1.3 % | 1419 ± 21 | 1.5 % |
|  | Proposed | 312 ± 4 | 1.3 % | 456 ± 5 | 1.1 % | 623 ± 6 | 1.0 % | 792 ± 6 | 0.8 % | 1151 ± 13 | 1.1 % | 1472 ± 21 | 1.4 % |
| Delay 4 s | Look-Locker | 290 ± 4 | 1.4 % | 432 ± 5 | 1.2 % | 603 ± 6 | 1.0 % | 768 ± 7 | 0.9 % | 1111 ± 12 | 1.1 % | 1386 ± 22 | 1.6 % |
|  | Proposed | 311 ± 4 | 1.3 % | 455 ± 5 | 1.1 % | 625 ± 6 | 1.1 % | 793 ± 6 | 0.8 % | 1151 ± 13 | 1.1 % | 1466 ± 23 | 1.6 % |
| Delay 3 s | Look-Locker | 290 ± 3 | 1.0 % | 432 ± 5 | 1.2 % | 600 ± 6 | 1.0 % | 760 ± 8 | 1.1 % | 1080 ± 11 | 1.0 % | 1325 ± 18 | 1.4 % |
|  | Proposed | 311 ± 4 | 1.3 % | 454 ± 5 | 1.1 % | 622 ± 8 | 1.3 % | 786 ± 8 | 1.0 % | 1143 ± 12 | 1.0 % | 1459 ± 25 | 1.7 % |
| Delay 2 s | Look-Locker | 289 ± 3 | 1.0 % | 430 ± 5 | 1.2 % | 592 ± 6 | 1.0 % | 740 ± 8 | 1.1 % | 1015 ± 13 | 1.3 % | 1215 ± 17 | 1.4 % |
|  | Proposed | 310 ± 4 | 1.3 % | 453 ± 5 | 1.1 % | 621 ± 7 | 1.1 % | 788 ± 9 | 1.1 % | 1142 ± 17 | 1.5 % | 1457 ± 33 | 2.3 % |
| Delay 1 s | Look-Locker | 286 ± 4 | 1.4 % | 412 ± 5 | 1.2 % | 549 ± 5 | 0.9 % | 667 ± 7 | 1.0 % | 779 ± 12 | 1.5 % | 989 ± 15 | 1.5 % |
|  | Proposed | 308 ± 4 | 1.3 % | 445 ± 5 | 1.1 % | 608 ± 9 | 1.5 % | 755 ± 9 | 1.2 % | 1123 ± 25 | 2.2 % | 1423 ± 65 | 4.6 % |
| IR Spin-Echo |  | 312 ± 6 | 1.9 % | 490 ± 8 | 1.6 % | 653 ± 6 | 0.9 % | 812 ± 8 | 1.0 % | 1175 ± 10 | 0.9 % | 1488 ± 16 | 1.1 % |

**Supporting Information Table S2.** $T_1$ relaxation times (ms, mean ± SD) and CoV values (%) for the experimental phantom in Figure 3.

|  | Tube 1 | | Tube 2 | | Tube 3 | | Tube 4 | | Tube 5 | | Tube 6 | |
|---|---|---|---|---|---|---|---|---|---|---|---|---|
|  | $T_1$ | CoV | $T_1$ | CoV | $T_1$ | CoV | $T_1$ | CoV | $T_1$ | CoV | $T_1$ | CoV |
| $\ell_1$-Wavelet (0.005) | 302 ± 8 | 2.6 % | 444 ± 13 | 2.9 % | 611 ± 15 | 2.5 % | 783 ± 21 | 2.7 % | 1148 ± 43 | 3.7% | 1457 ± 70 | 4.8% |
| $\ell_1$-Wavelet (0.02) | 303 ± 7 | 2.3 % | 444 ± 9 | 2.0 % | 611 ± 10 | 1.6% | 783 ± 13 | 1.7 % | 1144 ± 24 | 2.1% | 1454 ± 33 | 2.3% |
| $\ell_1$-Wavelet + Spatiotemporal TV (0.005) | 311 ± 5 | 1.6 % | 451 ± 4 | 1.0 % | 611 ± 4 | 0.7 % | 786 ± 4 | 0.5 % | 1144 ± 8 | 0.7% | 1454 ± 15 | 1.0% |
| IR Spin-Echo | 312 ± 6 | 1.9 % | 490 ± 8 | 1.6 % | 653 ± 6 | 0.9 % | 812 ± 8 | 1.0 % | 1175 ± 10 | 0.9% | 1488 ± 16 | 1.0% |

## II Oscillation Removal for Self-Gating

As described in [28], for the steady-state AC signal $\boldsymbol{X}$, we want to remove an oscillation which can be described by the basis:

$$n^t = \begin{pmatrix} e^{+i\psi_0 \cdot t} \\ e^{-i\psi_0 \cdot t} \\ e^{+2i\psi_0 \cdot t} \\ e^{-2i\psi_0 \cdot t} \\ \vdots \\ e^{+N_H i\psi_0 \cdot t} \\ e^{-N_H i\psi_0 \cdot t} \end{pmatrix},$$

with $\psi_0 \cdot t$ the projection angle at time $t$, $\psi_0$ the incremental projection angle and $N_H$ the highest order of the harmonic in the basis. Therefore, the corrected signal is $\boldsymbol{X}_{\mathrm{cor}} = \boldsymbol{X} - \boldsymbol{n}(\boldsymbol{n}^\dagger \boldsymbol{X})$ with $\dagger$ denoting the pseudo-inverse. $\boldsymbol{n}(\boldsymbol{n}^\dagger \boldsymbol{X})$ is the projection of $\boldsymbol{X}$ onto the basis $\boldsymbol{n}$.

For inversion prepared acquisitions, the contrast changes along time. The new centered k-space signal $\tilde{\boldsymbol{X}}$ can then be modeled by a global modulation for each channel $c$, i.e., $\tilde{X}_c(t) = \phi_c(t) X_c(t)$, where $\phi_c(t)$ describes the contrast change, and $\tilde{X}_c(t)$, $X_c(t)$ are the $c$th components of $\tilde{\boldsymbol{X}}$ and $\boldsymbol{X}$, respectively. As $X_c(t) = \sum_i n_i w_i + \eta$ with $w_i$ the weighting factor and $\eta$ the noise [28], $\tilde{X}_c(t) = \phi_c(t) X_c(t) = \sum_i \phi_c(t) n_i w_i + \phi_c(t) \eta$. Thus, our new basis for each channel becomes:

$$\tilde{\boldsymbol{n}}_c = \phi_c(t) \cdot \boldsymbol{n}_c$$

$\phi_c(t)$ can be determined using a moving average filter: $\phi_c(t) = \mathrm{movavg}(\tilde{X}_c(t))$. The newly corrected signal is then: $\tilde{\boldsymbol{X}}_{\mathrm{cor}} = \tilde{\boldsymbol{X}} - \tilde{\boldsymbol{n}}(\tilde{\boldsymbol{n}}^\dagger \tilde{\boldsymbol{X}})$.

## III Study of High-Dimensional Regularization on Simulated Dynamic Phantoms

To study the effects of various regularization types (explained in the Iterative Reconstruction part of the Methods section) in the motion-resolved reconstruction in general, a numerical motion phantom consisting of three elliptical objects was simulated. The diameters of the centered two objects were designed to

3

change along time, mimicking the cardiac contraction motion, moving from diastolic to systolic phase. Moreover, three small tubes, representing certain "lesions", are added to the "myocardium" region. The corresponding k-space data was then derived from the analytical Fourier representation of ellipses assuming an array of eight circular receiver coils surrounding the phantom without overlap. The simulation employed a continuous radial FLASH acquisition with a tiny golden-angle ($\approx 23,63°$) between successive spokes, base resolution of 256 pixels covering a field of view of 192 mm, 15 spokes per frame and a total of 32 time frames.

The effects of different regularizations were tested on the simulated dynamic phantom using motion-resolved image reconstruction ("PICS" command) in BART. The Supporting Information Figure S2 presents such a comparison where all reconstructions have employed the same regularization parameter. Certain streaking artifacts appear in the images when only the $\ell_1$-Wavelet-based spatial regularization is applied. Moreover, the reconstructed images are blurred with signal inhomogeneities on the myocardium regions. The above artifacts can be largely removed with the use of temporal TV regularization. However, the pure TV regularization also introduced the "line"-like artifacts along the time dimension, as shown in the line profile images. The joint $\ell_1$-Wavelet and temporal TV regularization helps to reduce such artifacts but still has some "line"-like artifacts presented in the line profiles. The spatiotemporal TV regularization, which takes account of both spatial and temporal information into the TV regularization, plus the $\ell_1$-Wavelet spatial regularization largely eliminate all the above artifacts, resulting in both smooth images and line profiles.

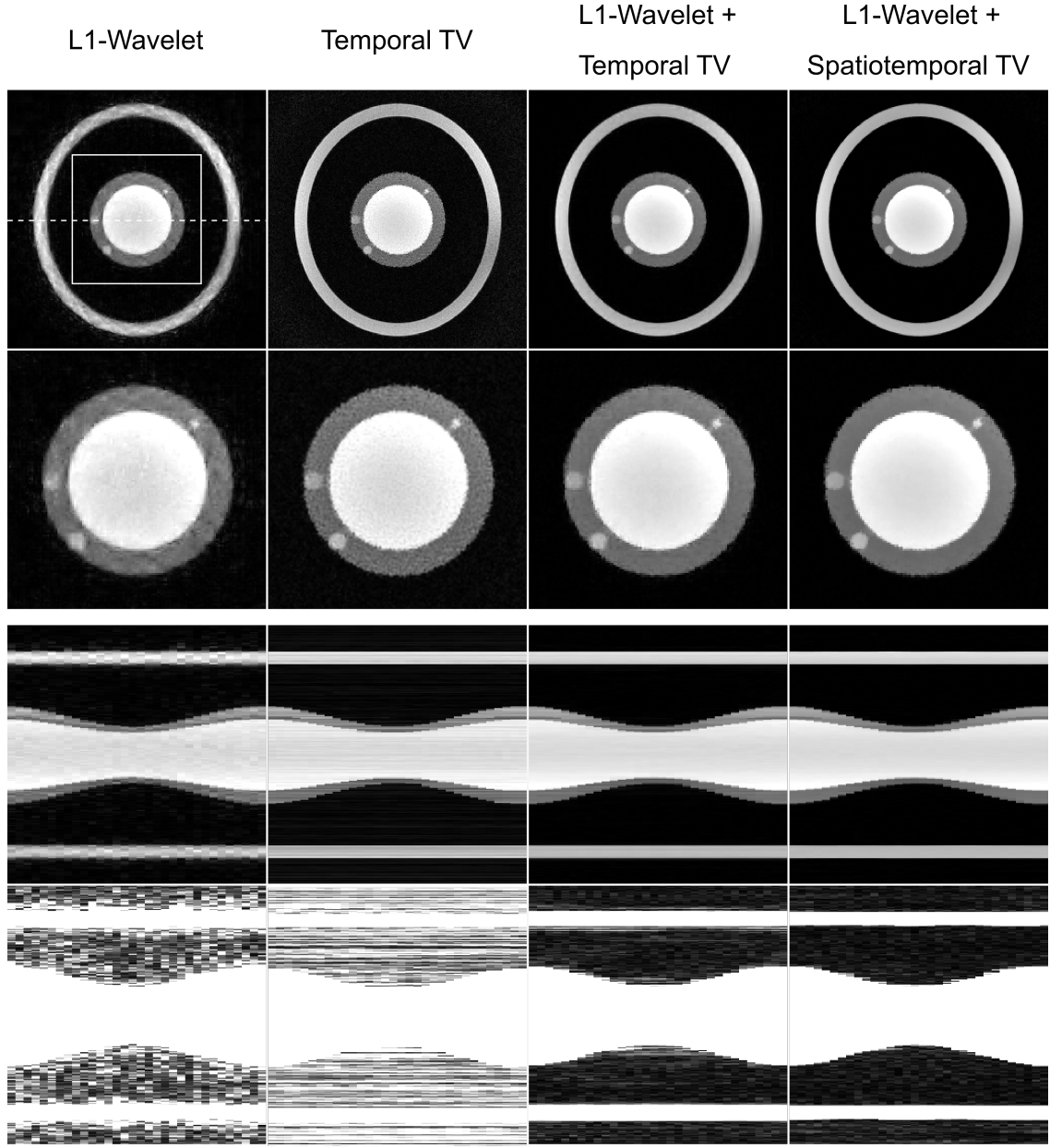

**Supporting Information Figure S2.** (Top) One of the "end-diastolic" images (28th of 32) of a simulated dynamic image series and (middle and bottom) the corresponding horizontal profiles (dashed line in the top) through the motion dimension. The image series were reconstructed using the motion-resolved image reconstruction with different regularization type.

## IV   Memory-efficient GPU Implementation

In general, the numerical optimization requires a huge amount of memory, involving the storage of the gridded k-space $Y$, the derivatives $\mathrm{D}F$, and temporary memory to hold intermediate results in during the computation of $F$ and $\mathrm{D}F$. The gridded k-space $Y$ and correspondingly $b$ require the most memory. Both have dimensions $(2N_x) \times (2N_y) \times N_C \times N_S \times T_{\boldsymbol{C}} \times T_{\boldsymbol{R}}$. With a typical problem size of $640 \times 640 \times 6 \times 27 \times 20 \times 3$ single precision complex float numbers, this corresponds to a memory of around 32GB. In contrast, the derivatives $\mathrm{D}F_{r,c}$ require about 10GB, the variables in the parameter space $(x, y, z)$ require about 2GB each and the temporary variables to compute the proximal operators require about 25GB in total. To

5

optimally utilize the available GPU-memory, we store the large variables $Y$ and $b$ in CPU-memory. The linear subproblem is solved completely on the GPU, however, all computations involving $F$, $\mathrm{D}F$ and $A^H A$ are performed independently and sequentially for the different motion states on the GPU. Thus, the complete gridded k-space does not need to be stored simultaneously on the GPU, as it would be the case if $A^H A$ was computed jointly for all motion states. With the above strategies, we are able to reduce the required GPU memory to about 70GB, making the model-based reconstruction feasible on a single NVIDIA A100 GPU with 80GB memory. Compared to a CPU (40-core 2.3 GHz Intel Xeon E5–2650 server with a RAM size of 512 GB) reconstruction, this strategy reduces the computational time from 12 hours to around 25 minutes. Moreover, the block-wise computation of $F$ and $\mathrm{D}F$ enables the distribution of the computations to multiple GPUs, allowing for further accelerations. For example, with a two-GPU system, the memory required on the first GPU reduces to about 60GB and the reconstruction time can be reduced to be about 15 minutes per dataset. Note that the required memory is not halved by using two GPUs since the computation of the proximal operators cannot be distributed.

# V  Supporting Information for In Vivo Studies

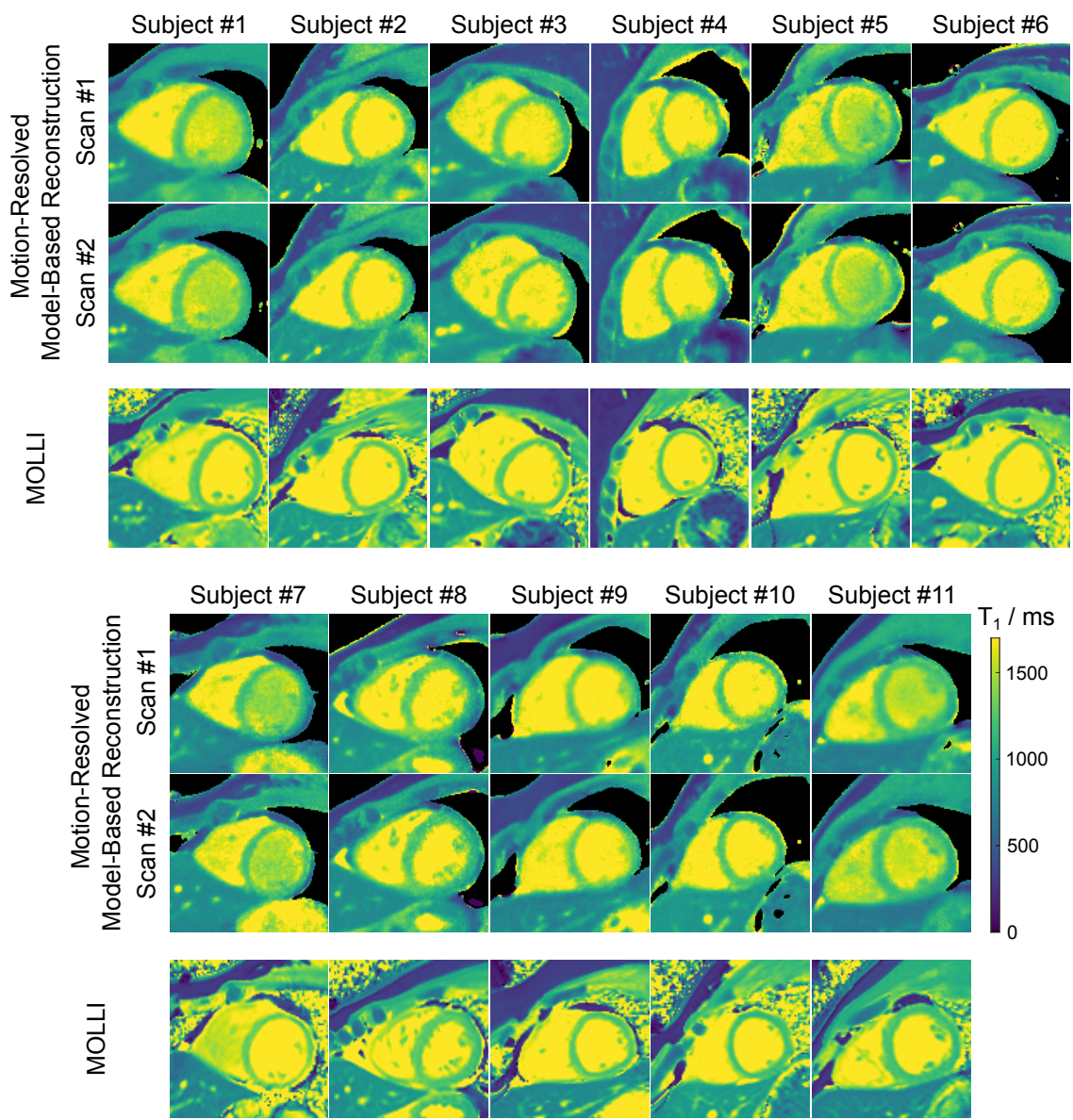

**Supporting Information Figure S3.** Two repeated myocardial $T_1$ maps (end-expiration, end-diastolic) of the motion-resolved model-based reconstruction acquired during free breathing in comparison to MOLLI acquired in a breathhold for all eleven subjects.

**Supporting Information Table S3.** The P-value of paired t-test comparison of the proposed motion-resolved model-based reconstructed diastolic $T_1$ values and the MOLLI values in each AHA segment for all scans.

|         | Anterior | Septal       |             | Inferior | Lateral       |              |
|---------|----------|--------------|-------------|----------|---------------|--------------|
|         |          | Anteroseptal | Inferoseptal |          | Inferolateral | Anterolateral |
| Basal*  | 0.99     | 0.58         | 0.85        | 0.56     | 0.07          | 0.25         |
| Middle  | 0.44     | 0.17         | 0.32        | 0.55     | < 0.01        | < 0.01       |
| Apical  | 0.48     | 0.27         |             | 0.89     | 0.69          |              |

* Basal and apical results are from six subjects, while the mid-ventricular results are evaluated on all eleven subjects.

**Supporting Information Table S4.** The relative difference (%) of the left-ventricular area between end-diastolic and end-systolic mid-ventricular myocardial $T_1$ maps for all subjects and scans.

| Subject | #1 | #2 | #3 | #4 | #5 | #6 | #7 | #8 | #9 | #10 | #11 | mean ± SD |
|---|---|---|---|---|---|---|---|---|---|---|---|---|
| Scan #1 | 57 | 57 | 58 | 68 | 65 | 57 | 68 | 65 | 63 | 65 | 54 | 62 ± 5 |
| Scan #2 | 60 | 54 | 59 | 65 | 62 | 54 | 67 | 57 | 68 | 65 | 55 | 61 ± 5 |

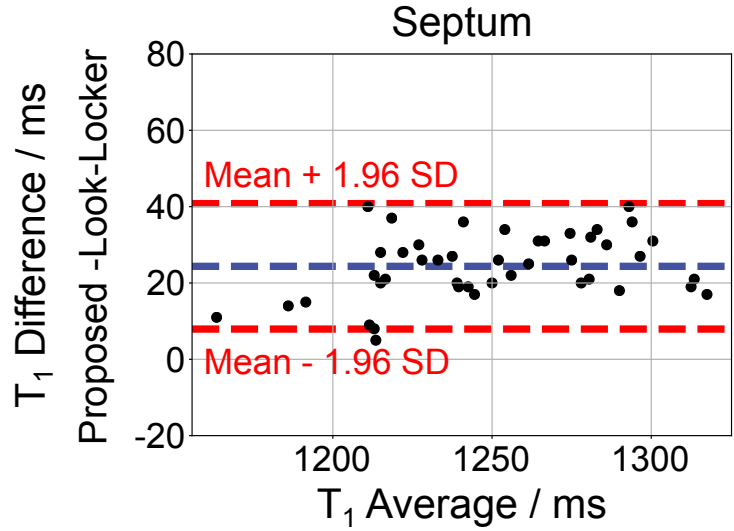

**Supporting Information Figure S4.** Bland-Altman plots comparing ROI-analyzed mid-ventricular septal mean $T_1$ values between the proposed correction and the Look-Locker correction for all subjects and scans. The mean difference is 24 ms and SD is 8 ms.

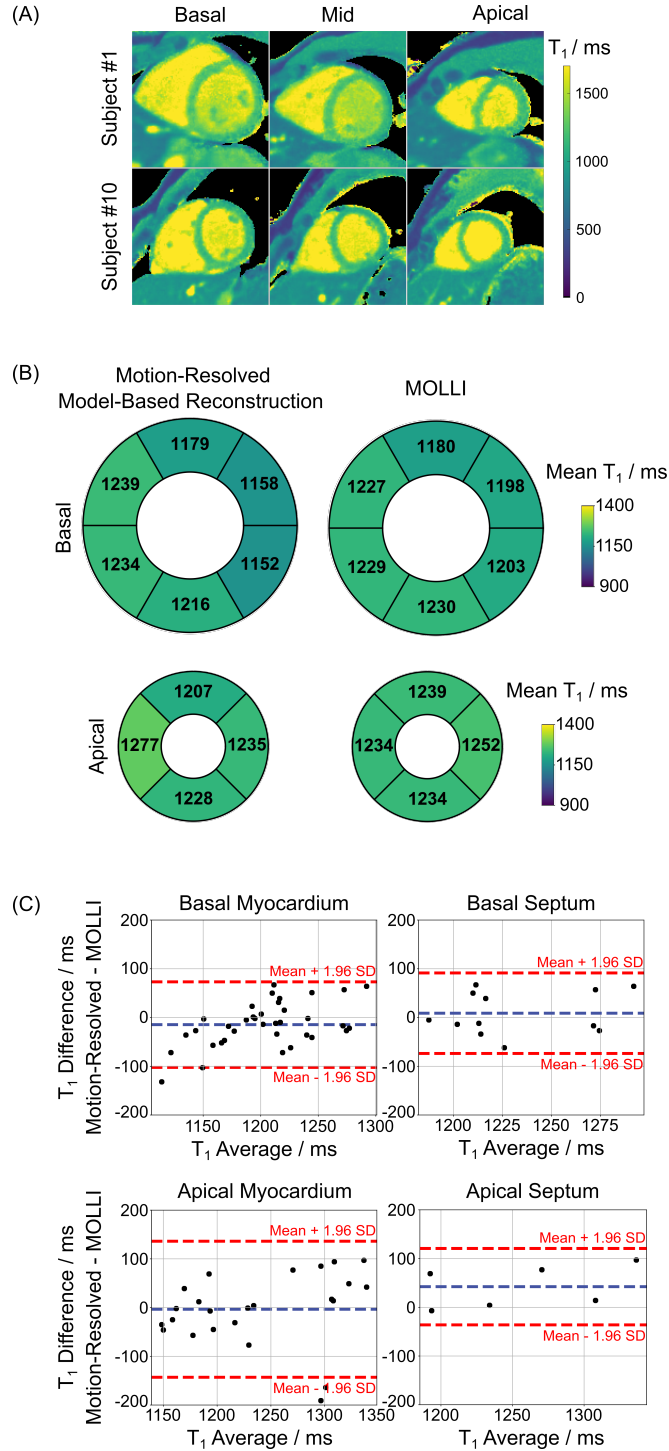

**Supporting Information Figure S5.** A. Representative diastolic $T_1$ maps of basal, mid and apical short-axis slices reconstructed with the motion-resolved model-based reconstruction (free breathing) for two healthy subjects. B. Bullseye plots of (top) six basal segments and (bottom) four apical segments, showing the mean diastolic $T_1$ values for six subjects for (left) the motion-resolved model-based reconstruction acquired during free breathing and (right) the MOLLI reference acquired in a breathhold. C. Bland–Altman plots comparing (top) the mean diastolic $T_1$ values of all six basal segments (mean difference: -15 ms, SD: 45 ms) and the two basal septal segments (segments 2 and 3 according to AHA, mean difference: 9 ms, SD: 42 ms), and (bottom) the mean diastolic $T_1$ values of all apical segments (mean difference: -3 ms, SD: 71 ms) and the one apical septal segment (segment 14 according to AHA, mean difference: 42 ms, SD: 40 ms) for the proposed method and MOLLI for six subjects.

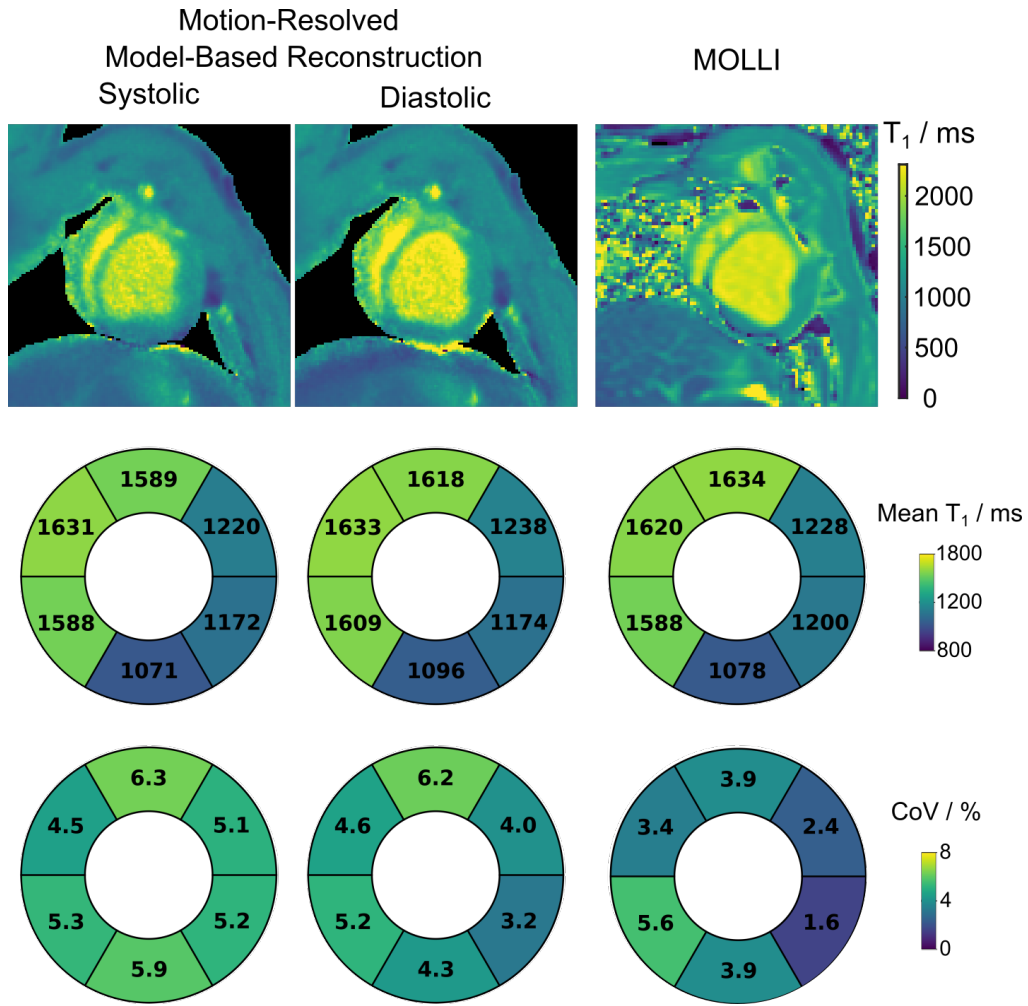

**Supporting Information Figure S6.** Free-breathing systolic and diastolic myocardial $T_1$ maps reconstructed with the motion-resolved model-based reconstruction and their comparison to the MOLLI reference for a pig study. The ROI-analyzed mean $T_1$ values (ms) and CoV values (%) for all the mid-ventricular AHA segments are presented in the bottom.


\textbf{Supporting Information Video~S1.} Synthesized $T_{1}$-weighted image series (bright blood and dark blood) and the corresponding myocardial $T_{1}$ maps through the cardiac phase dimension for subject $\#$2.

\textbf{Supporting Information Video~S2.} Synthesized $T_{1}$-weighted image series (bright blood and dark blood) and the corresponding myocardial $T_{1}$ maps through the cardiac phase dimension for subject $\#$3.

\textbf{Supporting Information Video~S3.} Synthesized $T_{1}$-weighted image series at temporal resolution of 49 ms for all inversion times and all cardiac phases of subject $\#$2.

\end{document}